\documentclass[useAMS,usenatbib]{mnras}

\usepackage{arydshln}
\usepackage{graphicx}
\usepackage{subfig}
\usepackage{natbib}
\usepackage{ulem}
\usepackage{subfig}
\usepackage{float}
\usepackage{multirow,multicol}
\usepackage{footnote}
\usepackage[bottom]{footmisc}
\usepackage{ulem}
\usepackage{xltabular}
\usepackage{longtable}
\usepackage{lipsum} 
\usepackage{pdflscape}
\usepackage{hyperref}
\usepackage{xcolor, colortbl}
\usepackage{lastpage}
\usepackage{verbatim}
\usepackage{booktabs}
\usepackage[flushleft]{threeparttable}
\hypersetup{
colorlinks = true, 
urlcolor = magenta, 
linkcolor = blue, 
citecolor = blue 
}
\usepackage[all]{hypcap}
\usepackage{txfonts}

\newcommand\xmm{{\it XMM-Newton}}

\newcommand\chandra{{\it Chandra}}
\newcommand\rosat{{\it ROSAT}}

\newcommand\asts{{\it AstroSat}}

\newcommand{\eqp}{EQ~Peg}

\newcommand{\E}[1]{$\times~10^{#1}$}
\newcommand{\Pten}[1]{$10^{#1}$}

\newcommand\ergs{$\rm{erg}~\rm{s}^{-1}$}       
\newcommand\arcm{\hbox{$^\prime$}}            
\newcommand\arcs{\hbox{$^{\prime\prime}$}}    
\newcommand\lx{$\rm{L}_{X}$}                   
\newcommand\lbol{$\rm{L}_{bol}$}                   
\newcommand\msun{$\rm{M}_{\odot}$}           
\newcommand\zsun{$\rm{Z}_{\odot}$}           
\newcommand\cts{counts s$^{-1}$}

\newcommand\apec{{\sc apec}}

\newcommand\chisq{$\chi^2$}

\newcommand\nh{$\rm{N_{H}}$}
\newcommand{\gsimeq}{\hbox{\raise0.5ex\hbox{$>\lower1.06ex\hbox{$\kern-0.92em{\sim}$}$}}}
\newcommand{\lsimeq}{\hbox{\raise0.4ex\hbox{$<\lower1.06ex\hbox{$\kern-0.92em{\sim}$}$}}}

\title[Superflares on EQ~Peg]{{\it AstroSat} observations of Long-duration X-ray superflares on active M-dwarf binary EQ~Peg}

\author[Karmakar et al.]{Subhajeet Karmakar\textsuperscript{1,2\thanks{E-mail: \href{mailto:subhajeet09@gmail.com}{subhajeet09@gmail.com, sk@mira.org}}}, Sachindra Naik\textsuperscript{2}, Jeewan C. Pandey\textsuperscript{3}, and Igor S. Savanov\textsuperscript{4}\\
\textsuperscript{1} Monterey Institute for Research in Astronomy (MIRA), 200 Eighth Street, Marina, California 93933, USA\\
\textsuperscript{2} Astronomy \& Astrophysics Division, Physical Research Laboratory, Navrangapura, Ahmedabad 380009, India\\
\textsuperscript{3} Aryabhatta Research Institute of observational sciencES (ARIES), Manora Peak, Nainital 263002, India\\
\textsuperscript{4} Institute of Astronomy, Russian Academy of Sciences, ul. Pyatniskaya 48, Moscow 119017, Russia\\
} 

\begin{document}
\date{Accepted 2021 October 21; Received 2021 October 21; in original form 2020 November 24}
\pagerange{\pageref{firstpage}--\pageref{LastPage}} \pubyear{2020}
\maketitle
\label{firstpage}

\begin{abstract}
We present a comprehensive study of three large long-duration flares detected on an active M-dwarf binary \eqp\ using the Soft X-Ray Telescope of the \asts\ observatory. 
The peak X-ray luminosities of the flares in the 0.3--7~keV band are found to be within $\sim$5--10~\E{30} \ergs. 
The e-folding rise- and decay-times of the flares are derived to be in the range of 3.4--11 and 1.6--24~ks, respectively. 
Spectral analysis indicates the presence of three temperature corona with the first two plasma temperatures remain constant during all the flares and the post-flare observation at $\sim$3 and $\sim$9~MK. The flare temperature peaked at 26, 16, and 17~MK, which are 2, 1.3, and 1.4 times more than the minimum value, respectively. The peak emission measures are found to be 3.9--7.1~\E{53}~cm$^{-3}$, whereas the abundances peaked at 0.16--0.26 times the solar abundances. 
Using quasi-static loop modelling, we derive loop-lengths for all the flares as 2.5$\pm$0.5~\E{11}, 2.0$\pm$0.5~\E{11}, and 2.5$\pm$0.9~\E{11}~cm, respectively. The density of the flaring plasma is estimated to be 4.2$\pm$0.8~\E{10}, 3.0$\pm$0.7~\E{10}, 2.2$\pm$0.8~\E{10}~cm$^{-3}$ for flares F1, F2, and F3, respectively. Whereas the magnetic field for all three flares is estimated to be $<$100 G. The estimated energies of all three flares are $\gsimeq$\Pten{34-35}~erg, putting them in a category of superflare. All three superflares are also found to be the longest duration flares ever observed on \eqp.
\end{abstract}


\section{Introduction}\label{sec:intro}

Flares on the Sun and stars are the most extreme evidence of magnetic activity in solar/stellar atmospheres. Our understanding of the flares is mostly developed on the basis of the Sun. Flares occur in close proximity to the active regions, which are the regions with localized magnetic fields. Magnetic loops from these active regions extend into the stellar corona. As the footpoints of these loops are jostled by the convective motions in the stars, they are twisted and distorted until magnetic reconnection occurs near the loop tops \citep{Parker-88-ApJ-21, Benz-10-ARA+A-2}. The reconnection process drives a rapid and transient release of magnetic energy in coronal layers, which is also associated with the electromagnetic radiation from radio waves to $\gamma$-rays. As a consequence, the charged particles are accelerated and gyrated downward along the magnetic field lines, producing synchrotron radio emission. These electron and proton beams then collide with the denser chromospheric materials, and hard X-rays ($>$20 keV) are emitted. Simultaneous heating of the plasma up to tens of MK evaporates the material from the chromospheric footpoints, which in turn increases the density of the newly formed coronal loops and emits at extreme UV and X-rays. Detailed and thorough investigations of these flaring events are useful to provide crucial information on the coronal structure. 

On the Sun and magnetically active stars, flares are found to be associated with mass loss due to coronal mass ejections \citep[CMEs;][]{GudelM-97-ApJ-2, DrakeJ-00-ApJ-1, YashiroS-06-ApJ-1}. In solar case, CMEs are observed to eject from 10$^{13-17}$~g of magnetized plasma into the interplanetary medium \citep[e.g.][]{YashiroS-09-IAUS, VourlidasA-10-ApJ-2}. Since most of the magnetically active cool stars can attain X-ray luminosities of about 1000 times more than that of the Sun, there is scope for vigorous CME activity, especially in the context of recent giant flare detections on solar-type stars \citep[][]{Maehara-12-Natur-2}. Typically the solar flares emit the energy of \Pten{29-32} erg within the flare duration of several minutes to several hours. Stellar flares  on solar-type stars with energies \Pten{33-38} erg are generally termed as `superflare' \citep{Schaefer-00-ApJ-10, Shibayama-13-ApJS-2}. Although there are thousands of superflares, have been observed to date in optical and UV band \citep{TuZ-20-ApJ}, but X-ray superflares are still very few \citep[see][and references therein]{Karmakar-17-ApJ-5}. 
It is to be noted that most of the host stars of the X-ray superflares are either an M-dwarf or a binary or multiple systems with an M-dwarf component. Moreover, M-dwarfs show a higher level of magnetic activities than other solar-type stars \citep[][]{Suarez-MascarenoA-16-A+A-2, KarmakarS-19-BSRSL}. This makes them very interesting objects to study.

M-dwarfs are the most populous low-mass stars with masses in the range of 0.6--0.075 \msun\ \citep{BaraffeI-96-ApJ-1}.  All the M-dwarfs seem to show different levels of magnetic activities. Active M dwarfs of spectral types M0--M4 are known to be strong coronal X-ray sources with X-ray luminosities often close to the saturation limit of L$_{X}/L_{bol} \sim $10$^{-3.3}$ \citep{FlemingT-93-ApJ-4, PizzolatoN-03-A+A-3}. These stars show frequent and strong flaring activities during which the X-ray luminosity increases by more than two orders of magnitude \citep[e.g.][]{Favata-00-A+A-3, GudelM-02-ApJ-5, SchmittJ-08-A+A-5}. On the other hand, late M-dwarfs with spectral types M6--M9 are very faint X-ray sources during their quiescence, although they produce transient X-ray luminosity enhancements by orders of magnitude during flares \citep[][]{RutledgeR-00-ApJ-6, SchmittJ-02-A+A-3, StelzerB-06-A+A-1}. 
The magnetic field generation and subsequent activities in M-dwarfs are closely tied to stellar rotation and age \citep[e.g.][]{Mohanty-03-ApJ-5, PizzolatoN-03-A+A-3, KiragaM-07-AcA}. The faster a star rotates, the stronger its magnetic heating and surface activity. Since the angular momentum loss from magnetized winds slows rotation in M-dwarfs, magnetic activity also decreases with age.

The active M~dwarf binary \eqp\ (BD+19 5116, GJ 896) has a record of frequent and large flaring activities across the electromagnetic spectrum. With an age of 950~Myr \citep[][]{ParsamyanE-95-Ap-1}, the visual binary system \eqp\ consists of an M3.5 primary (\eqp~A) and an M4.5 secondary (\eqp~B), separated by an angular separation of 5.$\!$\arcs8 \citep[][]{Liefke-08-A+A-2}. From \textit{Gaia} DR2 observations, \eqp\ is found to be located at a distance of 6.260$\pm$0.003 pc \citep[][]{Bailer-Jones-18-AJ-6}. With V magnitudes of 10.35 and 12.4, both the primary and secondary components are well known optical flare stars \citep[][]{Pettersen-76-OslR, Lacy-76-ApJS}. Microwave emission during quiescence was observed by \cite{JacksonP-89-A+A-1}, which they attributed to the brighter A component. \cite{ZborilM-98-MNRAS-2} estimated a rotational velocity of 14~km~s$^{-1}$ for EQ~Peg~A, while \cite{DelfosseX-98-A+A-2} analyzed the B component and derived the rotational velocity of 24.2$\pm$1.4~km~s$^{-1}$; both $v$~sin~$i$ values are rather high and thus consistent with a short rotation period system. Using SuperWASP transit survey data, \cite{NortonA-07-A+A-4} derived a photometric period of 1.0664~day for the EQ~Peg system. The EQ~Peg system is a strong X-ray and extreme ultraviolet (EUV) source with a number of recorded flares. \cite{Pallavicini-90-A+A-6} reported two flares observed with {\it EXOSAT}. The first one, with an atypically shaped light curve, was observed by \cite{Haisch-87-A+A-2} in the context of a simultaneous {\it EXOSAT} and {\it International Ultraviolet Explorer (IUE)} campaign. A second large-amplitude flare was observed by \cite{PallaviciniR-86-LNP-4}. Another large flare was observed by \cite{KatsovaM-02-ASPC-1} with {\it ROSAT} satellite with simultaneous optical photometry. A first approach to separate the A and B components in X-rays was undertaken with {\it XMM-Newton} \citep[][]{RobradeJ-04-A+A}. Although the two stars show considerable overlap owing to the instrumental point spread function, it was possible to attribute about three-quarters of the overall X-ray flux to the A component. A subsequent detailed spectral analysis without resolving the binary has been performed by \citep[][]{Robrade-05-A+A-2}. The first X-ray observation that allowed an unambiguous spectral separation of the two binary components was done with {\it Chandra}/HETG  \citep{Liefke-08-A+A-2}. \cite{Morin-08-MNRAS-3} also found that the Zeeman Doppler Imaging map of EQ~Peg~A had a strong magnetic field spot of 0.8 kG, while EQ~Peg~B had a strong spot of 1.2 kG.

In this paper,  we have investigated three superflares that occurred on the M-dwarf binary \eqp\ observed with the first Indian multi-wavelength space observatory \asts. All three flares are remarkable in their flare duration and the X-ray energies. We have organized the paper as follows. Observations and the data reduction procedure are discussed in Section~\ref{sec:obs}. Analysis and results from X-ray timing and spectral analysis, along with the time-resolved spectroscopy, are presented in Section~\ref{sec:result}. Finally, in Section~\ref{sec:discussion}, we have discussed our results in light of the loop modelling, energetics, loop properties, CMEs, and magnetic field strengths and presented our conclusion.
\begin{table}

	\centering
	\caption{Observations log of EQ~Peg with {\it AstroSat}: ObsID A07\_094T01\_9000003248.} 
	\label{tab:obslog}
    \tabcolsep=0.24cm
	\begin{tabular}{ccccc}
      \hline\hline\\[-2mm]
    Orbit &	Exp. Time  &	Start Time                         & End Time )\\[-0mm]
    No.   &	(ks)            &    (UTC)                              & (UTC)\\[2mm]
    \hline \hline \\[-2mm]
    21972 &   1.086  &   2019-10-21   13:34:28  &   2019-10-21  14:47:58 \\
    21973 &   3.328  &   2019-10-21   14:07:57  &   2019-10-21  16:36:51 \\
    21974 &   3.582  &   2019-10-21   16:12:53  &   2019-10-21  18:18:31 \\
    21975 &   3.589  &   2019-10-21   17:47:35  &   2019-10-21  20:02:20 \\
    21976 &   3.114  &   2019-10-21   19:21:43  &   2019-10-21  21:45:29 \\
    21977 &   2.871  &   2019-10-21   21:09:35  &   2019-10-21  23:31:47 \\
    21978 &   2.707  &   2019-10-21   22:57:26  &   2019-10-22  01:18:18 \\
    21980 &   3.406  &   2019-10-22   00:51:47  &   2019-10-22  04:25:38 \\
    21981 &   1.595  &   2019-10-22   04:17:52  &   2019-10-22  06:11:53 \\
    21982 &   1.937  &   2019-10-22   05:48:01  &   2019-10-22  07:49:32 \\
    21984 &   2.184  &   2019-10-22   07:42:03  &   2019-10-22  09:58:17 \\
    21985 &   2.814  &   2019-10-22   09:08:04  &   2019-10-22  11:43:10 \\
    21986 &   2.524  &   2019-10-22   11:03:25  &   2019-10-22  13:27:29 \\
    21987 &   2.745  &   2019-10-22   13:14:53  &   2019-10-22  15:10:57 \\
    21988 &   3.404  &   2019-10-22   14:52:27  &   2019-10-22  16:59:24 \\
    21989 &   3.630  &   2019-10-22   16:28:49  &   2019-10-22  18:41:26 \\
    21990 &   3.416  &   2019-10-22   18:04:30  &   2019-10-22  20:23:13 \\
    21991 &   3.000  &   2019-10-22   19:56:00  &   2019-10-22  22:07:52 \\
    21992 &   3.000  &   2019-10-22   21:30:11  &   2019-10-22  23:55:50 \\
    21995 &   5.289  &   2019-10-22   23:22:36  &   2019-10-23  04:55:43 \\
    21996 &   0.404   &   2019-10-23  04:47:45  &   2019-10-23  05:50:52 \\[2mm]
	\hline
    \normalsize
	\end{tabular}


\end{table}

\begin{figure*}
\centering
\includegraphics[width=14cm, trim={1.2cm 0 3cm 0}, angle=0]{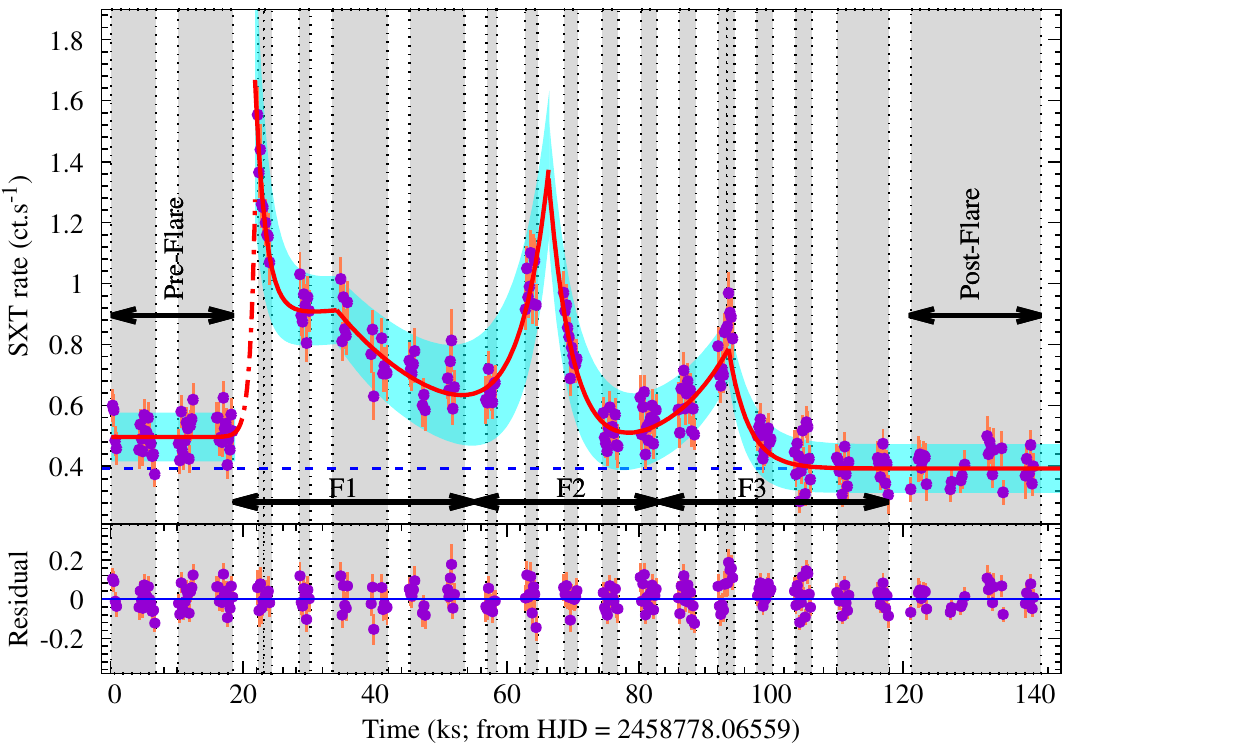}
\caption{In the top panel, the background-subtracted {\it AstroSat}/SXT light curve of EQ~Peg is shown in 0.3--7 keV energy band. The temporal binning of the light curve is 200~s. The solid red line shows the superimposed best-fit model to the light curve. While the cyan shaded region shows the 1$\sigma$ variations from the best-fit model. The grey shaded regions separated by the vertical dotted lines show the different time intervals for which time-resolved spectroscopy is performed. The approximate duration of the flares (F1, F2, and F3), the `Pre-flare', and the `Post-Flare' segments have been marked with black arrows. The red dot-dashed line is an extrapolated exponential curve that indicates the rise phase of flare F1. In the bottom pane, the (data -- model) residual have been plotted. The evenly distributed residuals along the horizontal line corresponding to zero residual indicate that the model curve does not seem to show any systematic deviation from the data.
}
\label{fig:lc}
\end{figure*}

\section{Observations and Data Reduction}\label{sec:obs}
We have observed \eqp\ using Soft X-ray focusing Telescope \citep[SXT;][]{SinghK-14-SPIE-1, SinghK-17-JApA-7}, Large Area X-ray Proportional Counter \citep[LAXPC;][]{AntiaH-17-ApJS}, Cadmium Zinc Telluride Imager \citep[CZTI;][]{BhaleraoV-17-JApA-3}, and Ultra-Violet Imaging Telescope \citep[UVIT;][]{TandonS-17-AJ-1}  onboard \asts\ observatory, during 2019 October 21--23 (PI. Karmakar; ID: A07\_094T01\_9000003248). We did not detect \eqp\ in hard X-ray during our observation, while the source was not observed with either near ultraviolet (NUV; 2000--3000 \AA) or far ultra-violet (FUV; 1300–1800 \AA) channels of UVIT. Since the visual (VIS; 3200--5500 \AA) channel is only meant for aspect correction and not expected for science observation \citep[][]{TandonS-20-AJ}, we have not used that in our analysis. In this paper, we have used the 0.3--7 keV \asts/SXT observations for further analysis. 
A detailed log of the observations of the source is given in Table~\ref{tab:obslog}. The level~1 data were processed using the {\tt AS1SXTLevel2-1.4b} pipeline software to produce level~2 clean event files for each orbit of observation. We used the {\tt sxtevtmerger}  tool\footnote{\href{https://www.tifr.res.in/~astrosat_sxt/dataanalysis.html}{https://www.tifr.res.in/$\sim$astrosat\_sxt/dataanalysis.html}} to merge the event files corresponding to each orbit to a single cleaned event file for further analysis. We extracted images, light curves, and spectra using the {\tt FTOOLS} task {\tt xselect} V2.4k, which has been provided as a part of the latest {\sc heasoft} version 6.28. For our analysis, we used a circular region of a radius of 13\arcm\ centred at the source position to extract source products, whereas multiple sources free circular regions of 2.$\!$\arcm6 radii have been chosen as background region. For spectral analysis, we used the spectral redistribution matrix file ({\tt sxt\_pc\_mat\_g0to12.rmf}), and ancillary response file ({\tt sxt\_pc\_excl00\_v04\_20190608.arf}) provided by the SXT team\footnote{\href{https://www.tifr.res.in/~astrosat_sxt/dataanalysis.html}{https://www.tifr.res.in/$\sim$astrosat\_sxt/dataanalysis.html}}. All the SXT spectra were binned to achieve a minimum of 10 counts per bin. The spectral analysis was carried out in an energy range of 0.3--7 keV using the X-ray spectral fitting package \citep[{\tt xspec};  version 12.11.1][]{Arnaud-96-ASPC-2}. The errors were estimated for a confidence interval of 68\% ($\Delta$\chisq\ = 1), equivalent to $\pm$1$\sigma$. In our analysis, the solar photospheric abundances (\zsun) were adopted from \cite{Anders-89-GeCoA-2}, whereas we used the cross-sections obtained by \cite{Morrison-83-ApJ-32} to model the equivalent hydrogen column density \nh.

\section{Analysis and Results}\label{sec:result}
\subsection{X-Ray Light Curves}\label{sec:lc}
The background-subtracted 0.3--7 keV X-ray light is shown in the top panel of Figure~\ref{fig:lc}. The {\it AstroSat}/SXT observation of \eqp\ began at HJD = 2458778.06559 day (=T0).
The light curve shows a large variation in SXT count rates from 0.4 to 1.6 \cts\ during the observation of $\sim$140 ks stare time. 
From the beginning of the observation until T0+20 ks, the light curve shows a constant count rate of $\sim$0.5 \cts, which we term as the `pre-flare' epoch. 
Whereas, a minimum count rate of $\sim$0.4 \cts\ has been observed during the observation from T0+120 ks to T0+140 ks, which we term as the `post-flare' epoch.

In this paper, in order to identify the flares, we have considered the positive count rate excursions greater than three times the standard deviation ($\sigma$) of the post-flare light curve.
It is to be noted that, Figure~\ref{fig:lc} shows an optimum light curve with a time binning of 0.2 ks, whereas median uncertainty in SXT count rate is $\sim$0.05 \cts. Therefore, using the \asts/SXT observation the best precision in estimating  $\tau_{r}$ and $\tau_{d}$ we can get is $\sim$3 min. On the other hand, since, to identify a flare, we need at least three consecutive points more than 3$\sigma$ level, we cannot identify any flare which lasts for $\lsimeq$ 10 minutes. Moreover, the timing characteristics of the {\it AstroSat}/SXT observations shows $\sim$45 minutes alternation between observation and data gap. This makes it more difficult to discriminate between long coherent flare profiles and a superposition of multiple shorter, smaller flares. 
From the \asts/SXT observation, we have identified three events (F1, F2, and F3) that can be significantly considered as flares. The identified flares have been shown with the black arrows in the top panel of Figure~\ref{fig:lc}. 

In order to model the light curve, we have fitted pre-flare and post-flare segments with a horizontal straight line. Whereas to model the flare light curve, prmarily we have fitted the rise and decay phase of the flares with the simple exponential rise and exponential decay, respectively, given by the following equation.
\begin{equation}
  \begin{array}{l}
  I(t)~=~I_{peak}.exp\left(\frac{t_{peak}~-~t}{\tau_{r,d}}\right)~+~I_{q} \\
  ~~~~~{\rm for~rise}~t \leq t_{peak}~~~;~~{\rm for~decay}~t \geq t_{peak}~
  \end{array}
\label{eq:trtd}
\end{equation}
\noindent
Where t$_{peak}$ is the flare peak time, $I_{peak}$ and $I_{q}$ are the flare peak count rate and minimum count rate, respectively. $\tau_{r}$ and $\tau_{d}$ are e-folding rise- and decay-time. I(t) is a time-dependent parameter that shows the count-rate variation during flare rise or decay. 
In order to get the best-fit parameters, we used the orthogonal distance regression technique considering uncertainties in both x-y directions.
The best-fit modelled light curve is shown in the top panel of Figure~\ref{fig:lc} using a solid red line, whereas the cyan shaded region shows the 1$\sigma$ variation of the best-fit model. The bottom panel shows the residual of the data from the best-fit model.  
The evenly distributed residuals along the horizontal line corresponding to zero residual indicate that the model curve does not seem to show any systematic deviation from the data.

Flares F1 started $\sim$T0+20 ks and extended until T0+58 ks, after reaching a peak count rate of $\sim$1.6 \cts. The rise phase of flare F1 could not be observed due to the data gap. In Figure~\ref{fig:lc}, we have shown this by an extrapolated exponential curve with a red dot-dashed line. During the decay phase, instead of a simple exponential decay, it seems to have a plateau phase around 34 ks. If we fit a function containing a simple exponential decay (for the flare F1) and an exponential rise (for flare F2) in the time interval T0+20 to T0+65 ks, we find a systematic deviation around 34 ks, which results in the reduced \chisq\ value of 1.4 for 58 degrees of freedom (DOF). However, instead of a simple exponential decay, if we fit two exponential decay where the second one peaks around 34 ks, we get a better fit with a reduced \chisq\ value of 0.9 for 56 DOF as shown in the top panel of Figure 1. This suggests that the decay of the flare F1 is best explained with a double-decay exponential. 
Either this might be due to a smaller flare superimposed to the flare F1, or the flare F1 is itself a double-decay flare. 
In order to verify if the plateau is likely to be a separate flare or not, we have detrended the data using the same function, but this time best-fit was obtained excluding the plateau time interval T0+25 to T0+42 ks. We find that the {\it AstroSat}/SXT counts during the plateau time interval varies within a 1-sigma uncertainty level. Therefore, we can not significantly identify this as a separate flare. Therefore, for further analysis, we assume this plateau is a part of flare F1; rather, flare F1 is considered as a double-decay flare. The e-folding decay times for flare F1 are estimated to be 1.6$\pm$0.4 and 24$\pm$5 ks. This kind of double decay flare have been observed on plenty of low mass stars, such as AU Mic \citep[][]{CullyS-93-ApJ-2},  UV Cet \citep[][]{BoppB-73-ApJ-2}, AD Leo \citep[][]{GershbergR-67-SvA-3}.

The flare F2 is found to start right after the flare F1 and ends around T0+84 ks after reaching a peak of $>$1.1 \cts. From manual inspection, flare F2 is more likely peaked during the data gap near 68 ks. The flare F3 started just after the flare F2 and lasts until $\sim$T0+120 ks while it reaches up to a peak count rate of $\sim$1 \cts. Both the flares F2 and F3 seems to show an exponential increase and exponential decay. 
From T0+68 to T0+93 ks, an exponential decay for flare F2 was fitted along with an exponential rise for flare F3 with best-fit reduced \chisq\ value of 1.14 for 53 DOF. The fourth piece was fitted from T0+93 to T0+120 ks as an exponential decay for flare F3 with a best-fit reduced \chisq\ value of 1.32 at 44 DOF. 
The e-folding rise times for flare F2 and F3 are estimated to be 3.4$\pm$0.8 and 11$\pm$2 ks, whereas the respective decay times are estimated to be 3.0$\pm$0.6 and 3.1$\pm$0.4 ks.
 This indicates that the rise and decay times in flares F2 are comparable in the soft X-ray band. However, flare F3 shows a comparatively slower rise than its decay.

\begin{figure*}
\centering
\subfloat[]{\includegraphics[height=6.0cm, angle=-90]{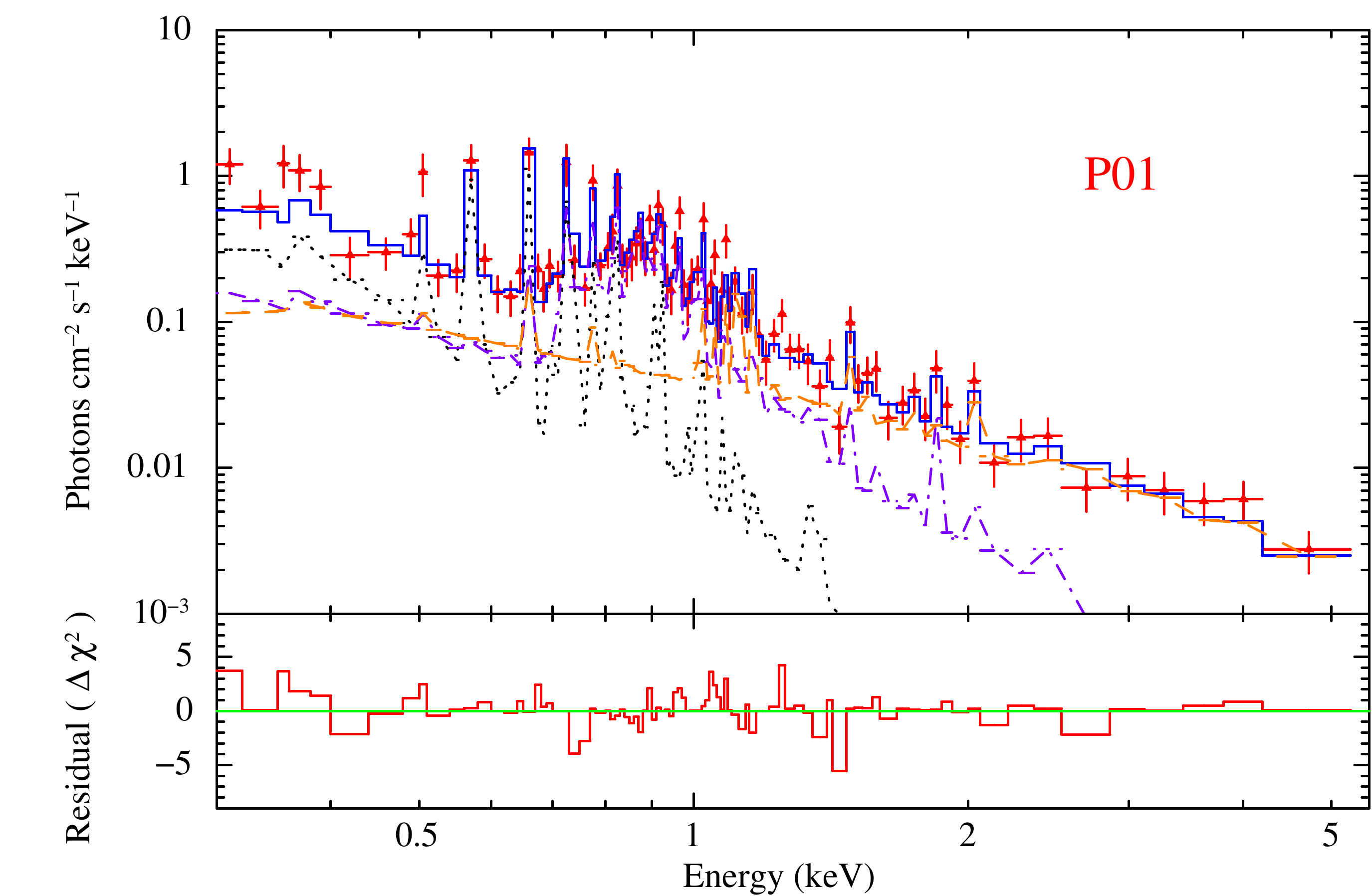}}
\subfloat[]{\includegraphics[height=6.0cm, angle=-90]{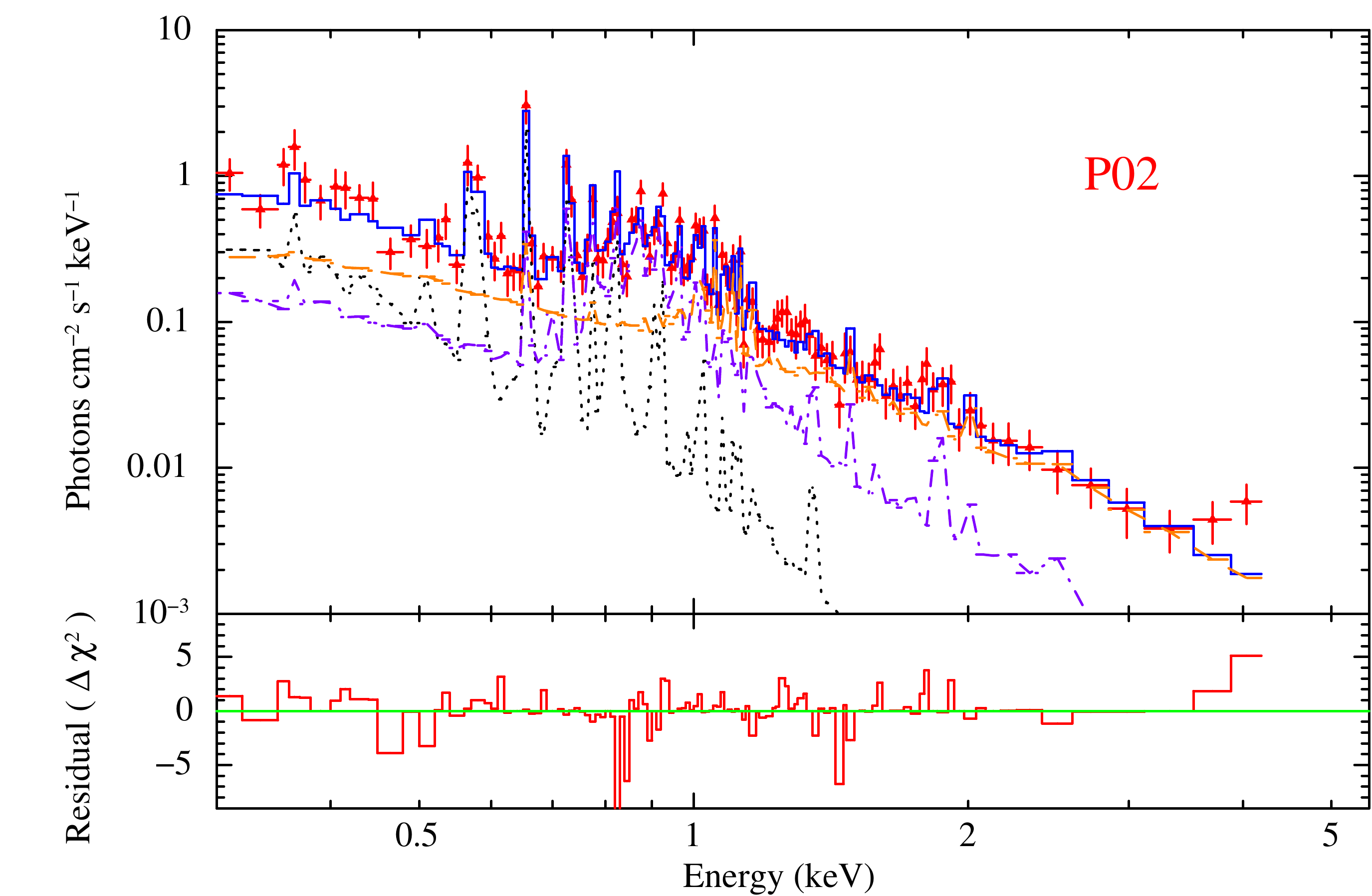}}
\subfloat[]{\includegraphics[height=6.0cm, angle=-90]{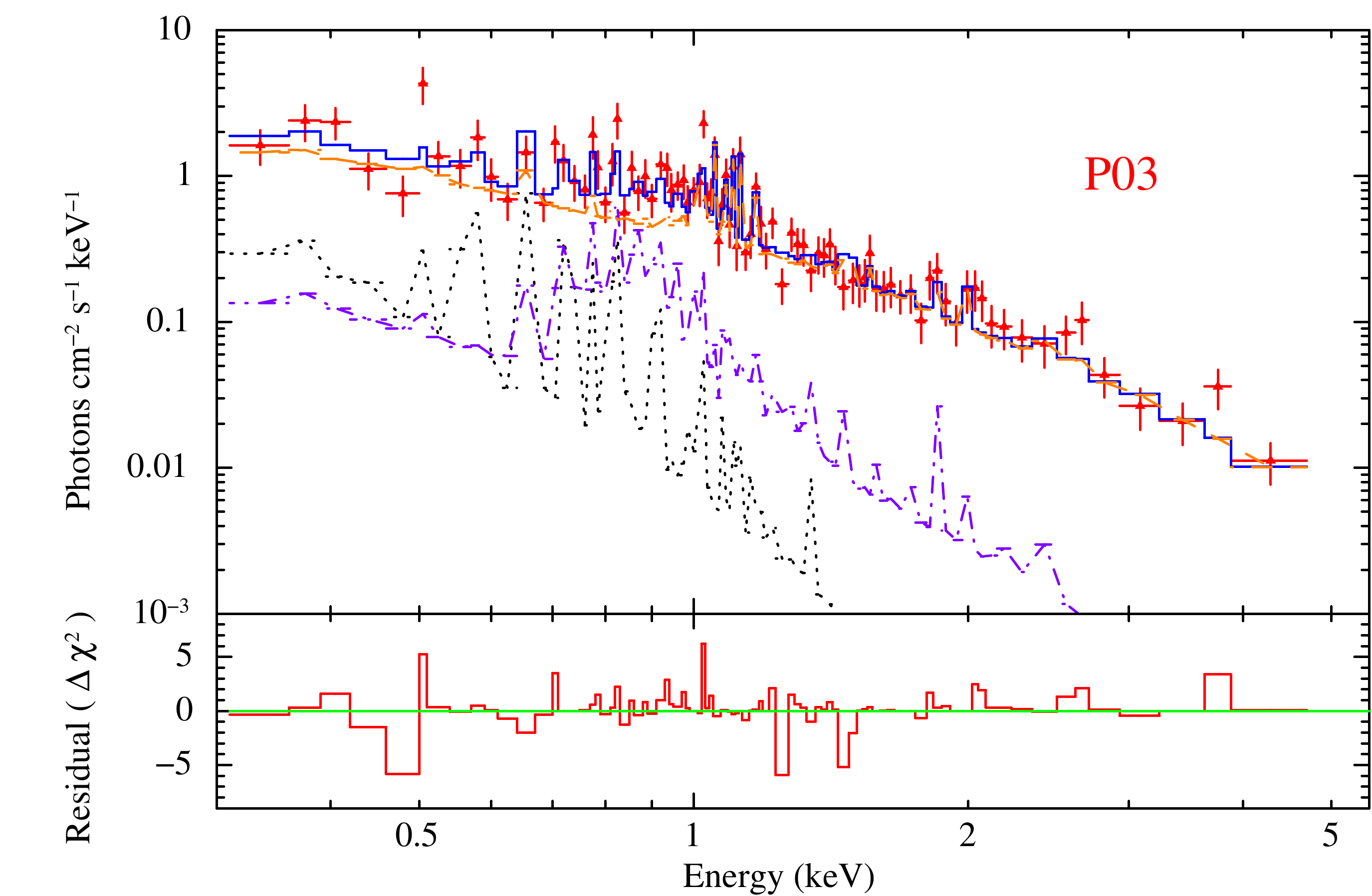}}
\\[-4.2mm]
\subfloat[]{\includegraphics[height=6.0cm, angle=-90]{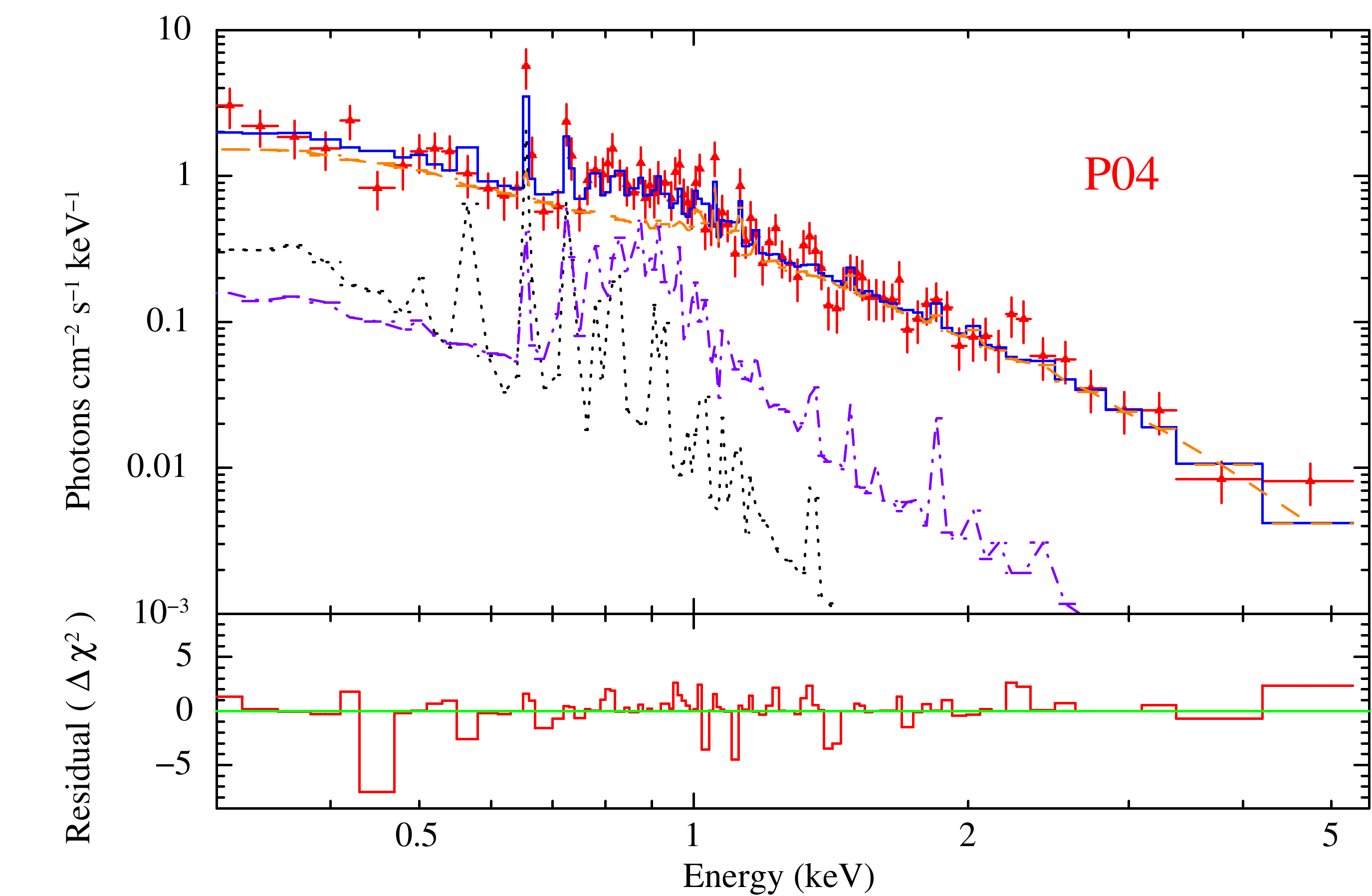}}
\subfloat[]{\includegraphics[height=6.0cm, angle=-90]{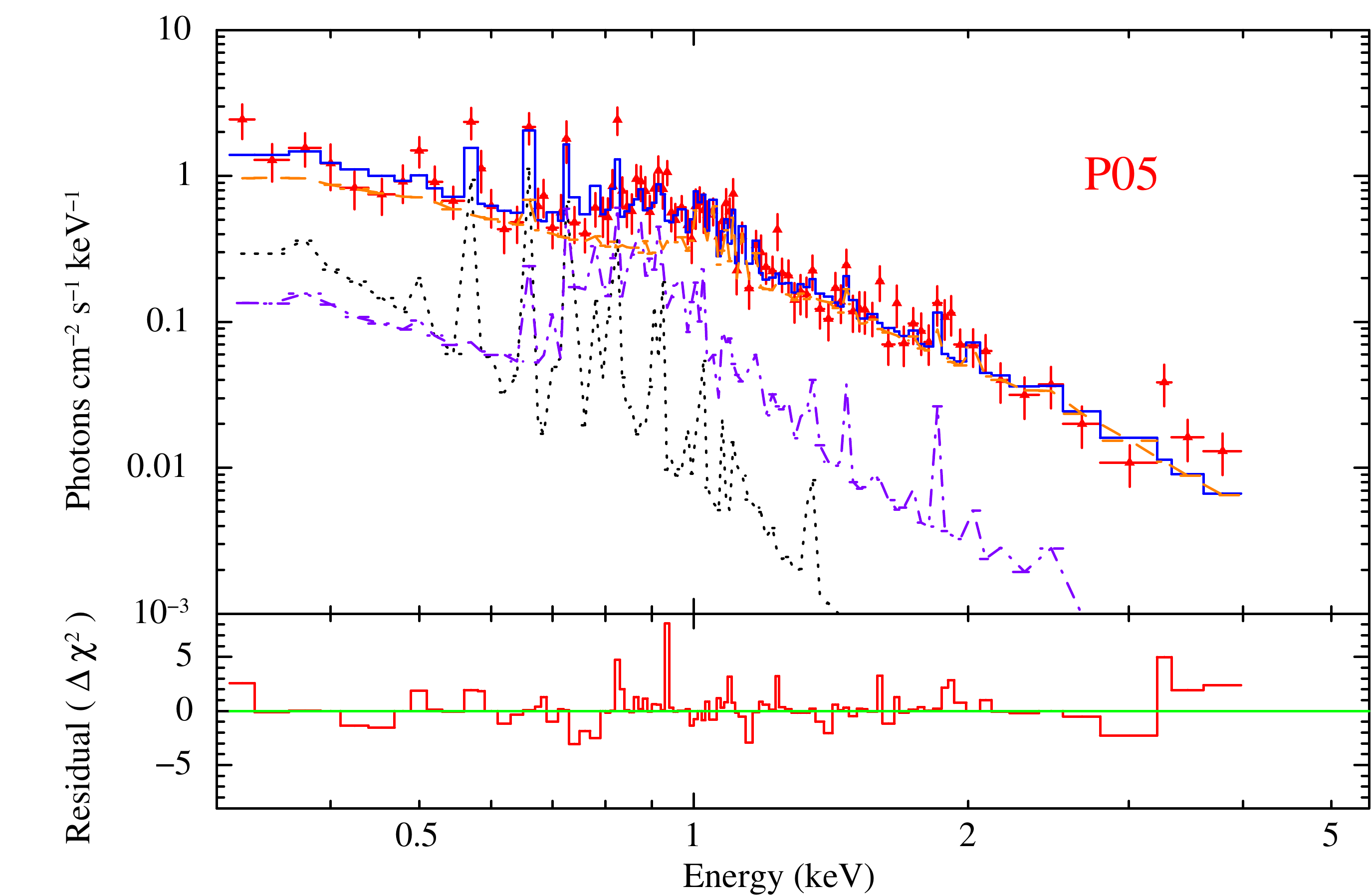}}
\subfloat[]{\includegraphics[height=6.0cm, angle=-90]{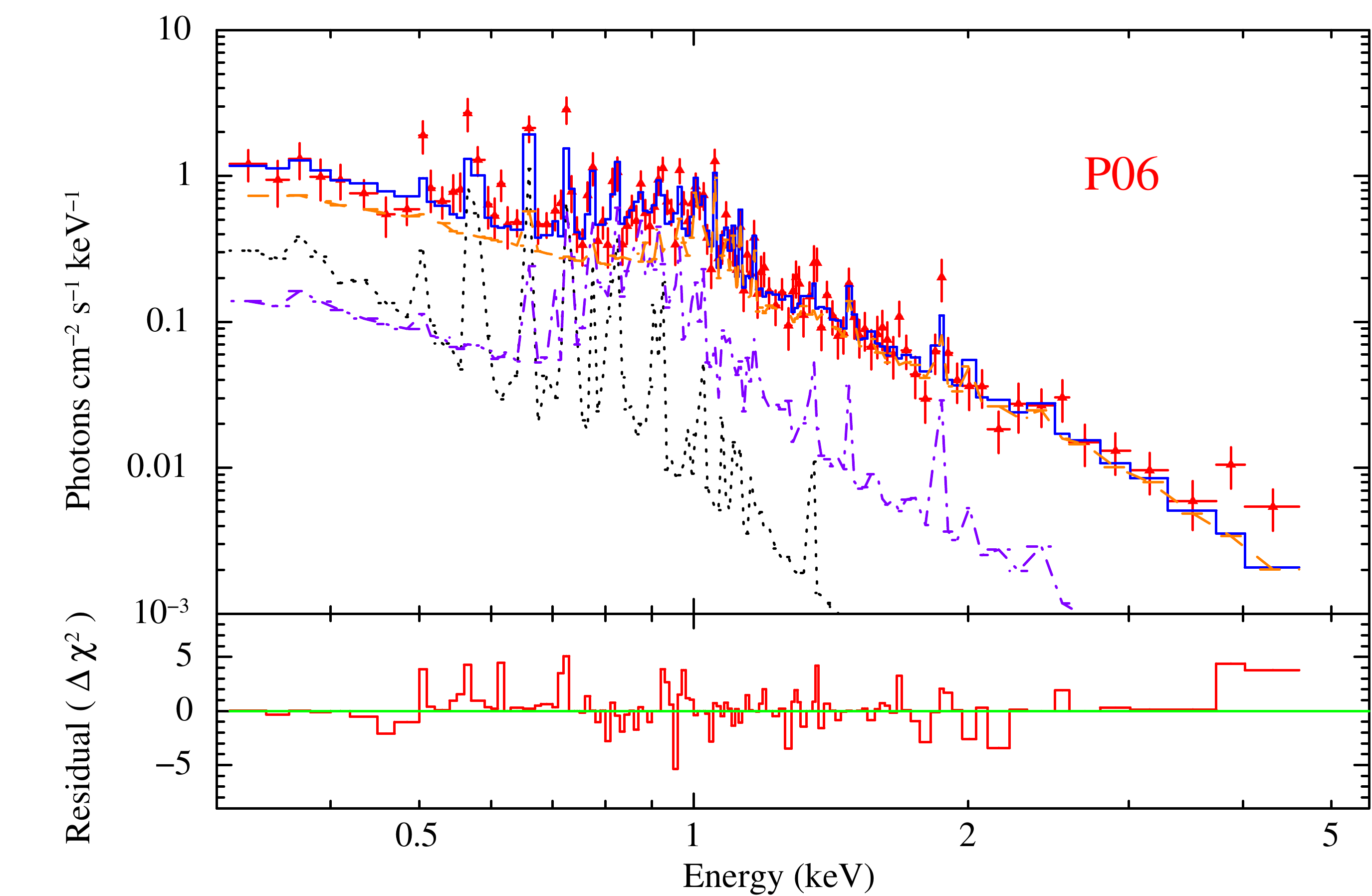}}
\\[-4.2mm]
\subfloat[]{\includegraphics[height=6.0cm, angle=-90]{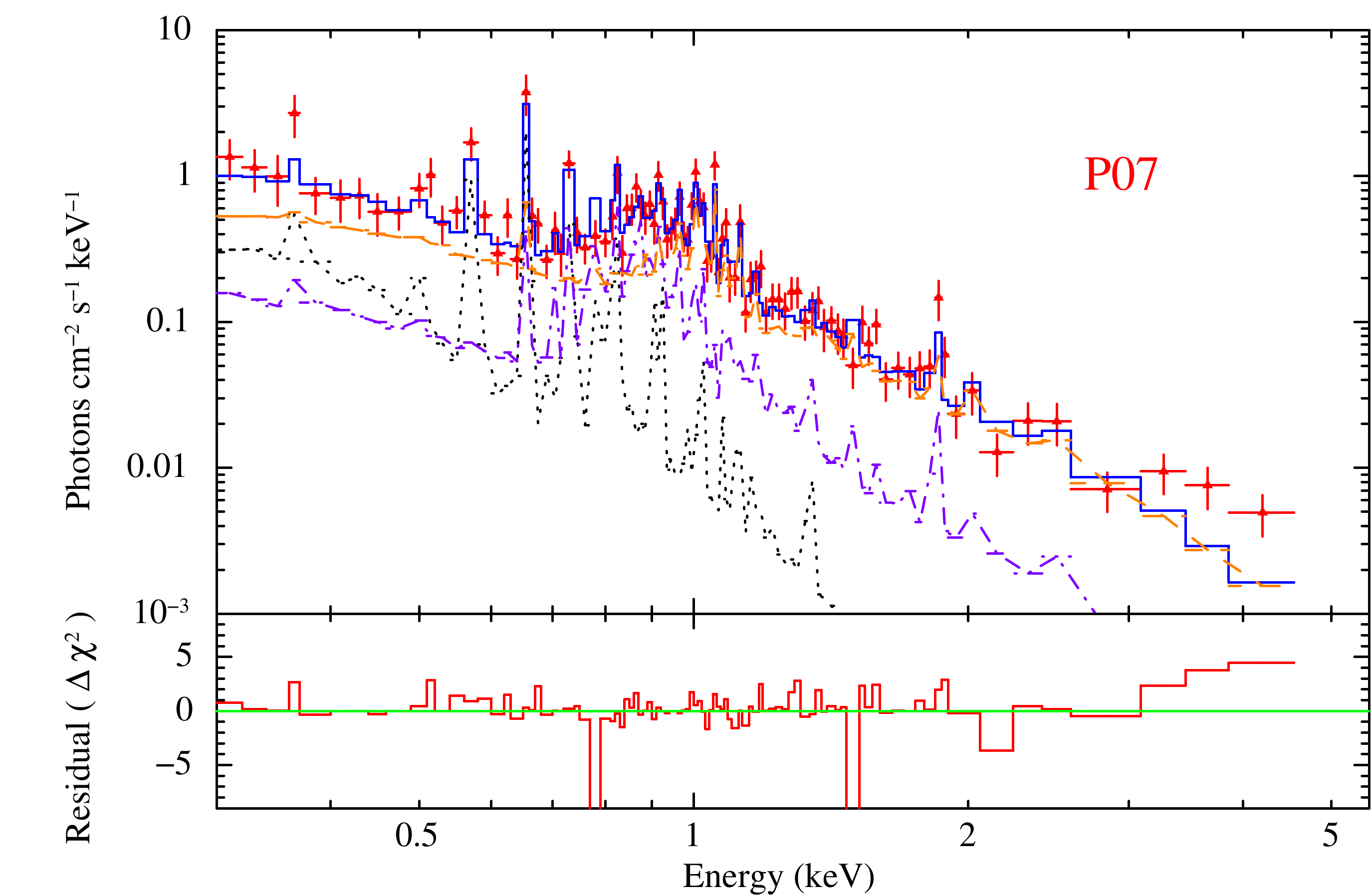}}
\subfloat[]{\includegraphics[height=6.0cm, angle=-90]{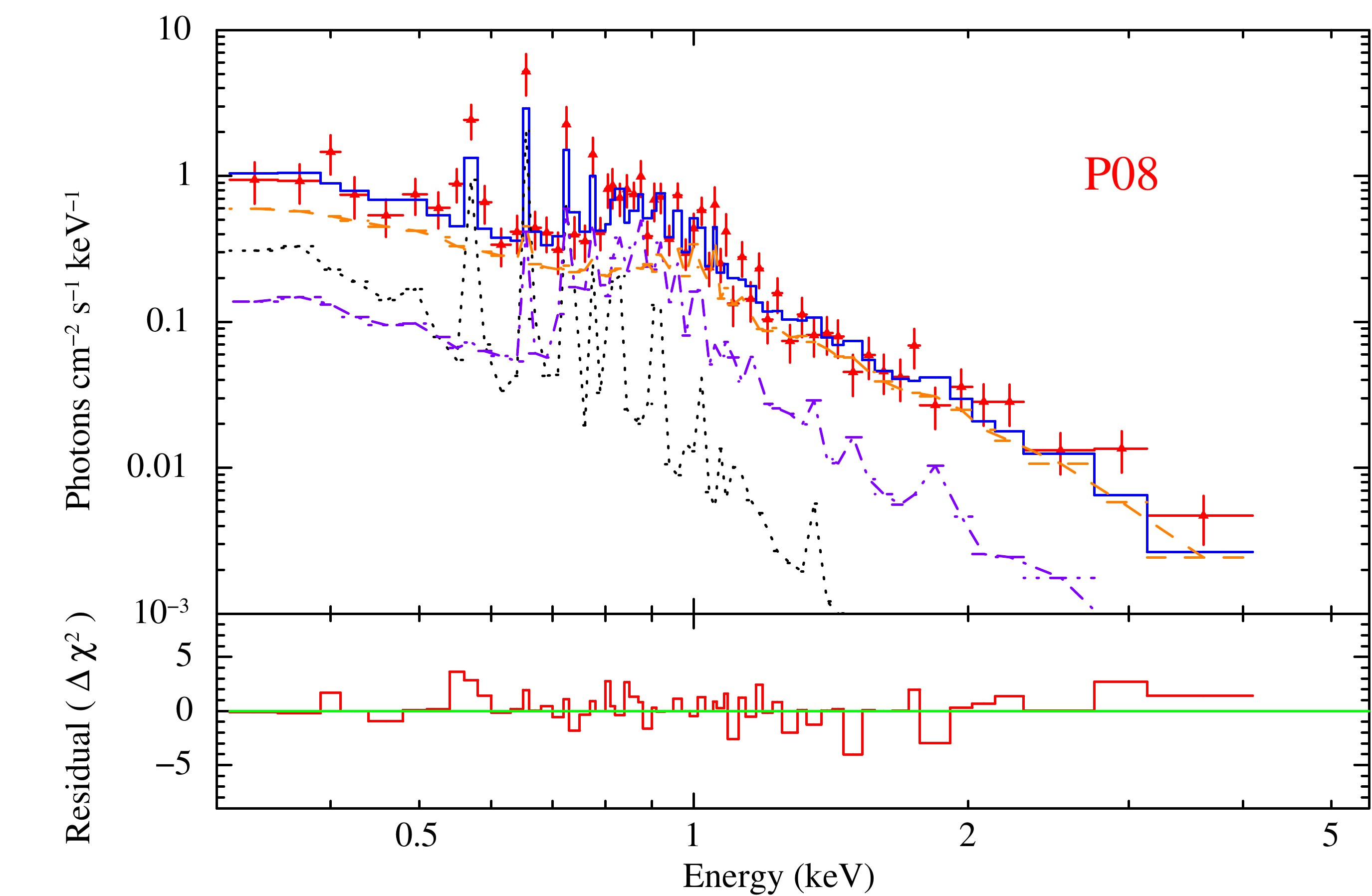}}
\subfloat[]{\includegraphics[height=6.0cm, angle=-90]{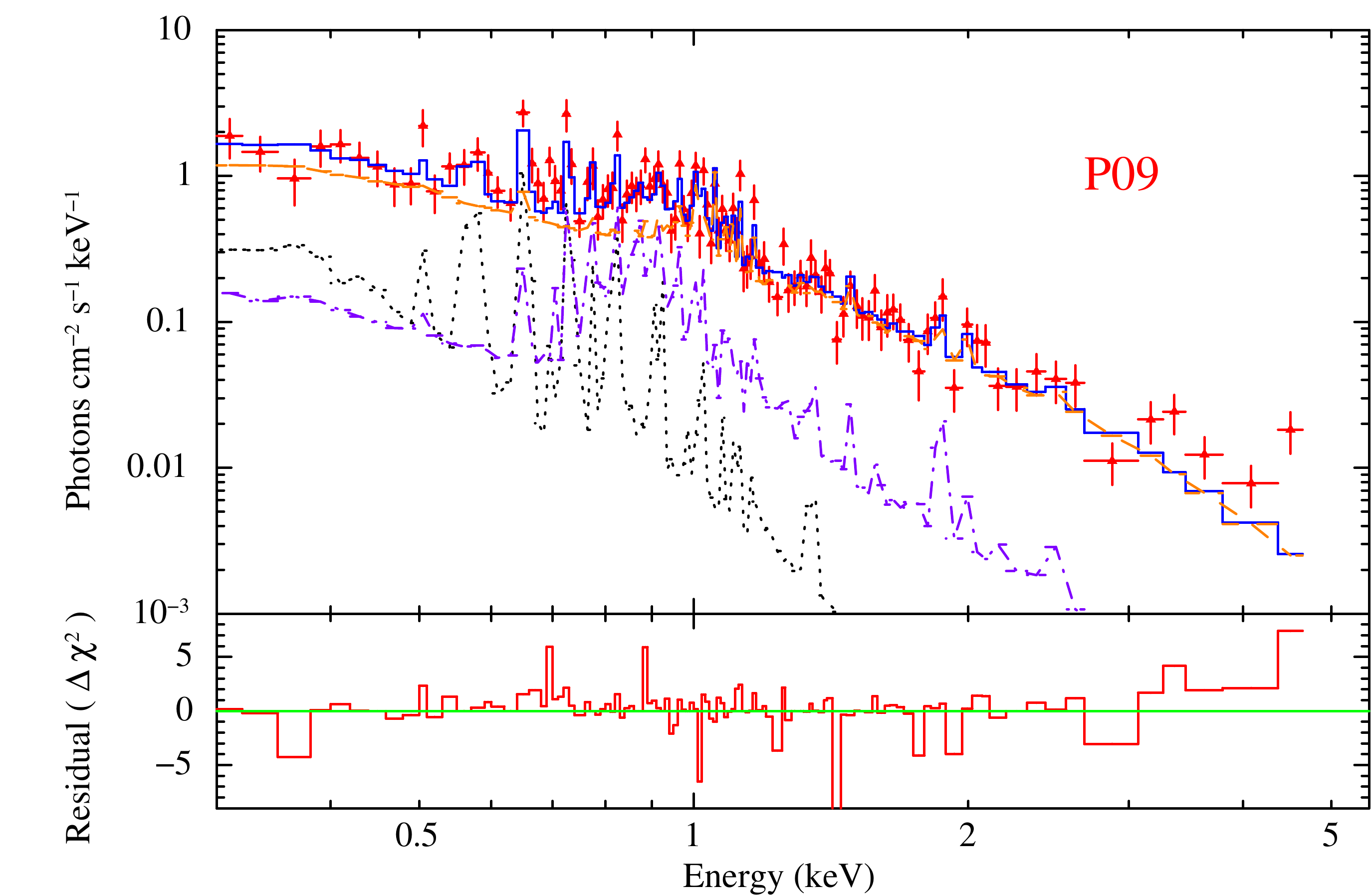}}
\\[-4mm]
\subfloat[]{\includegraphics[height=6.0cm, angle=-90]{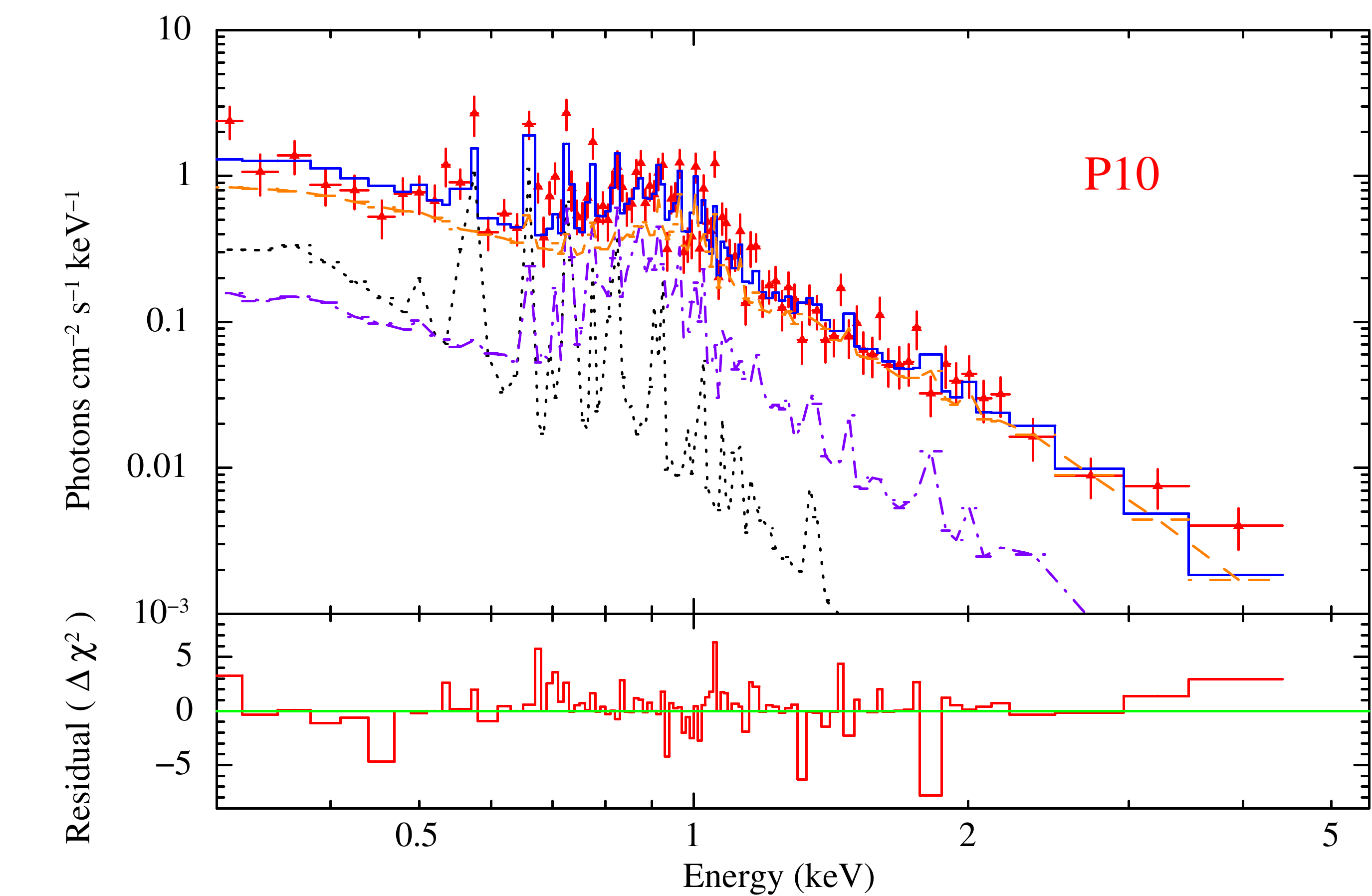}}
\subfloat[]{\includegraphics[height=6.0cm, angle=-90]{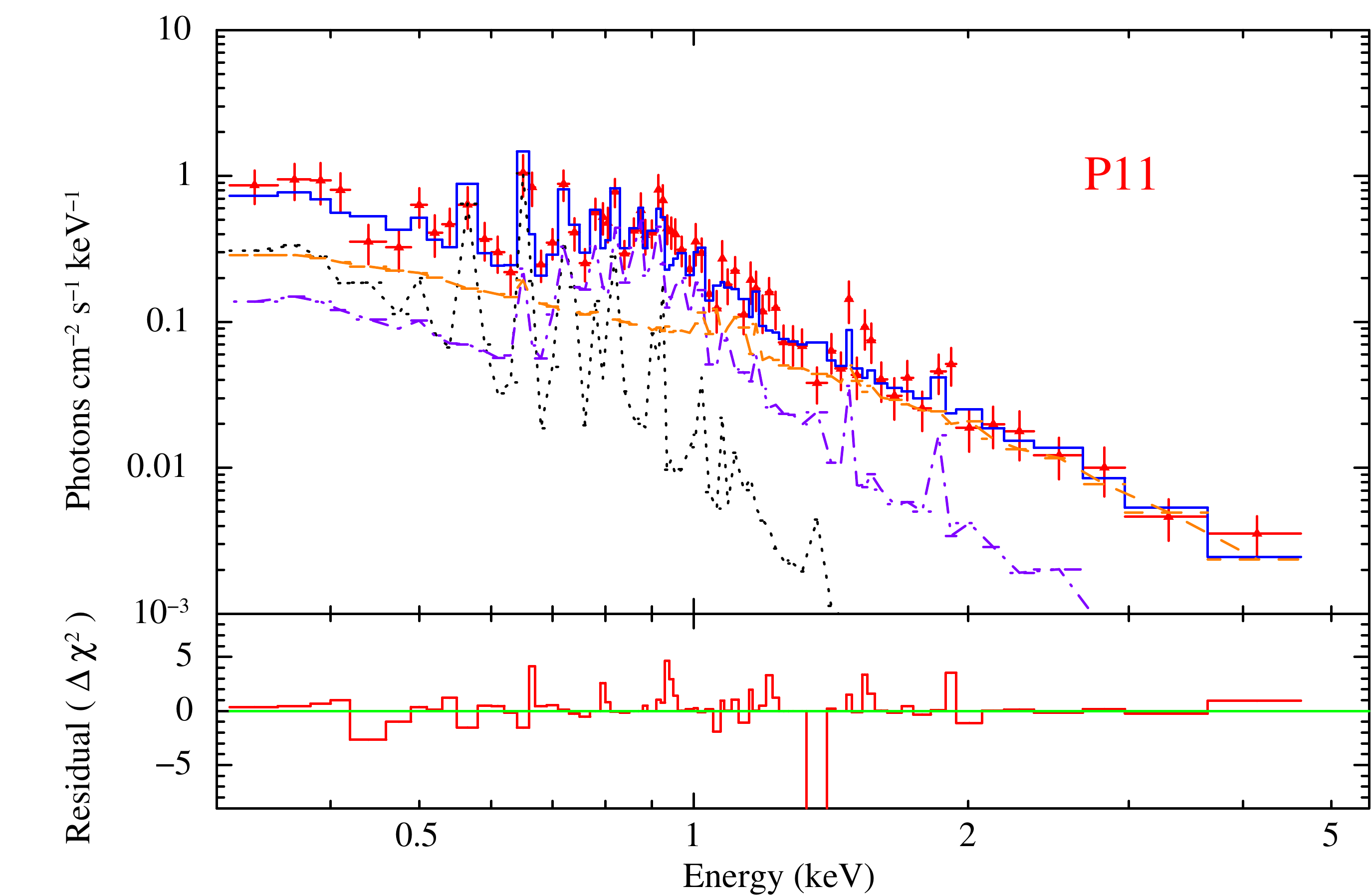}}
\subfloat[]{\includegraphics[height=6.0cm, angle=-90]{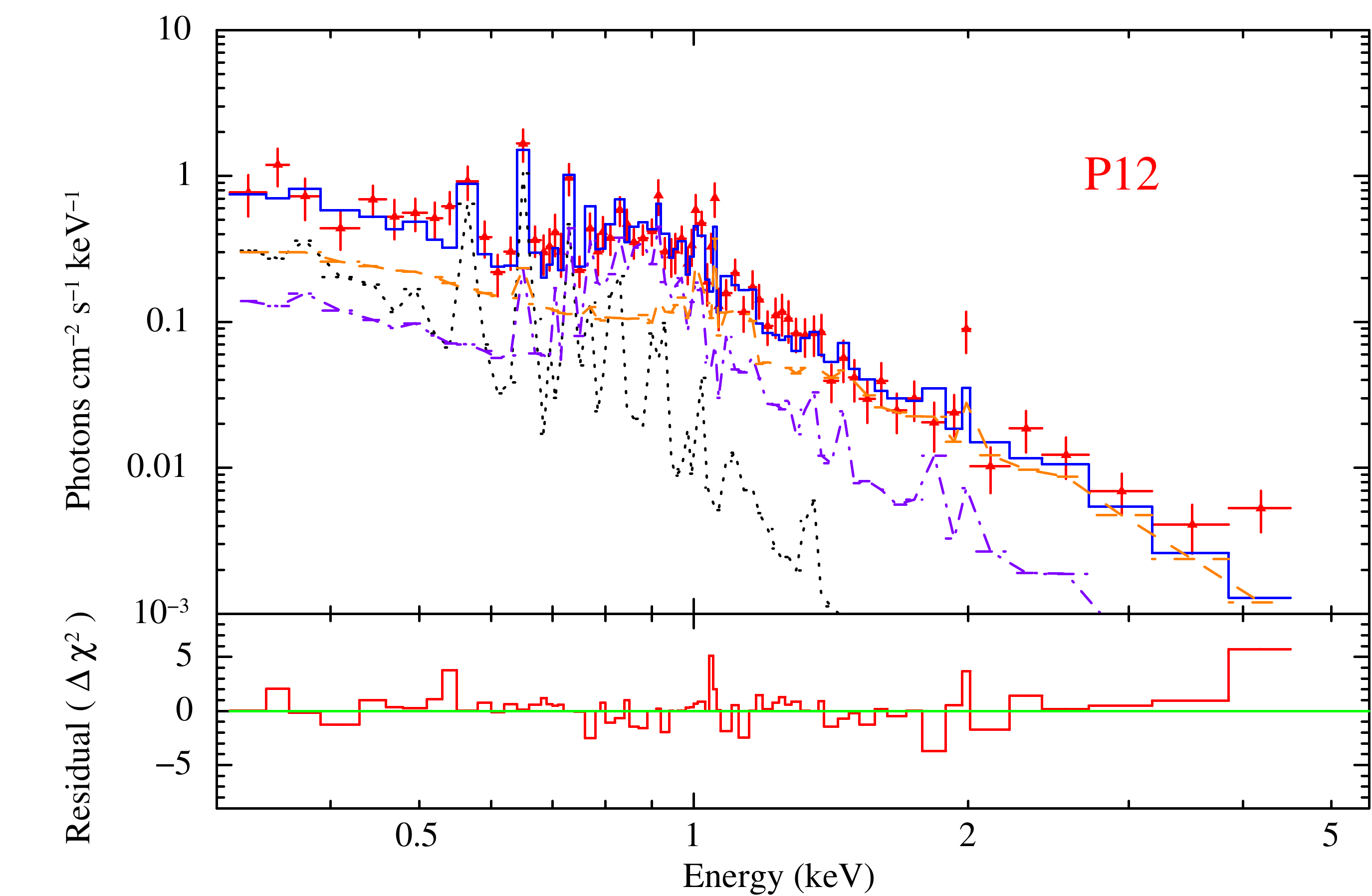}}
\\[-4.2mm]
\subfloat[]{\includegraphics[height=6.0cm, angle=-90]{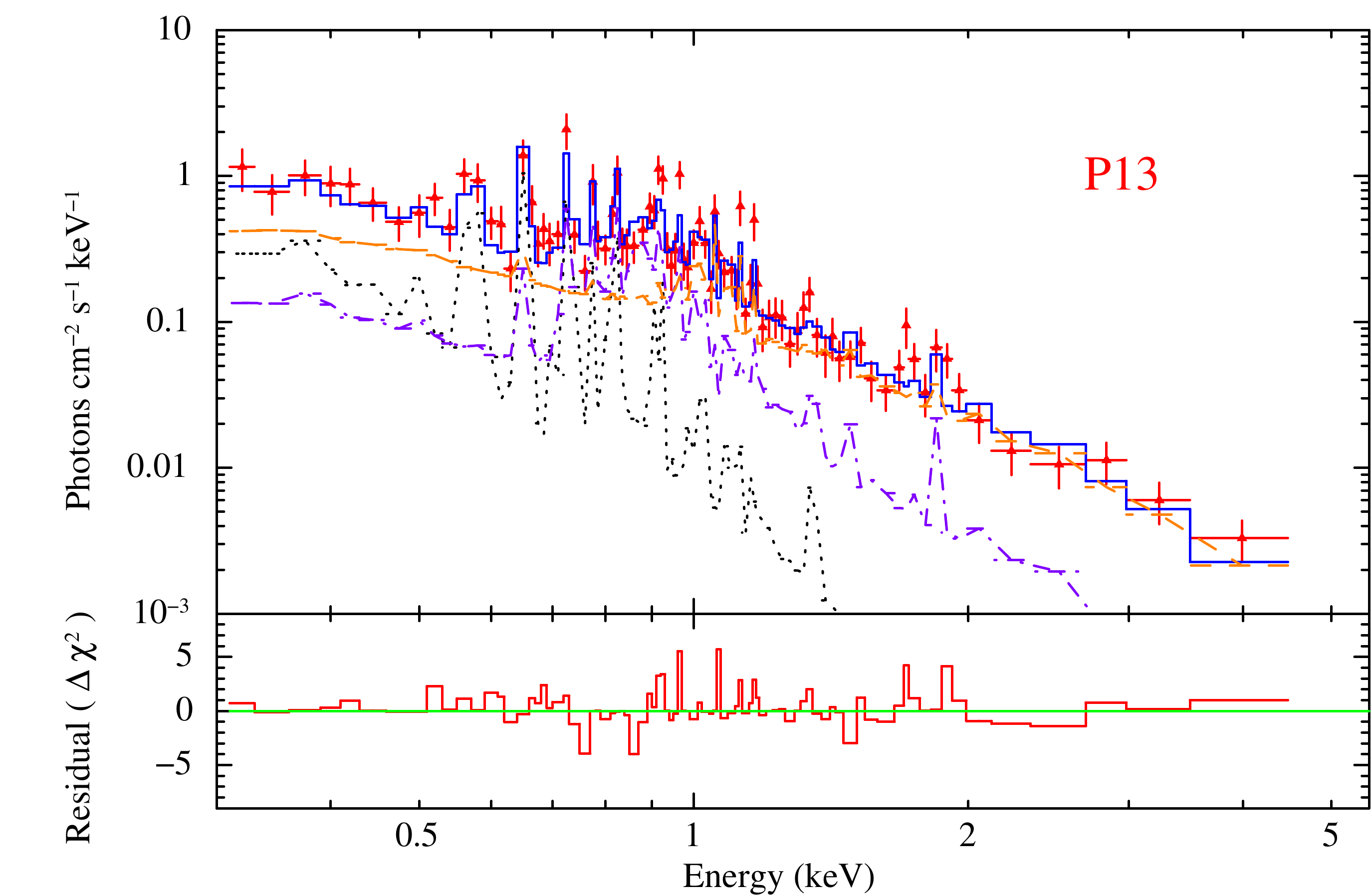}}
\subfloat[]{\includegraphics[height=6.0cm, angle=-90]{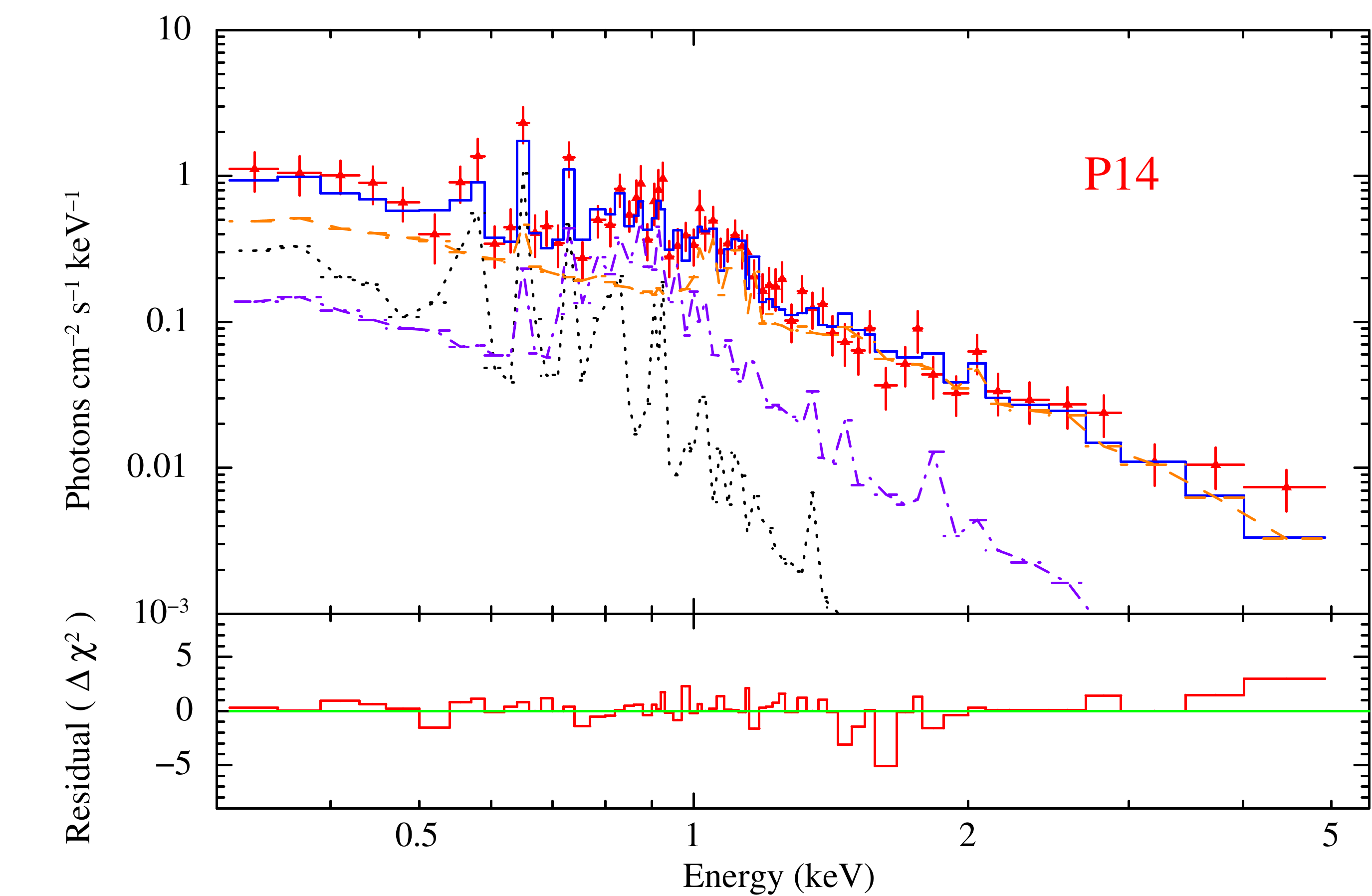}}
\subfloat[]{\includegraphics[height=6.0cm, angle=-90]{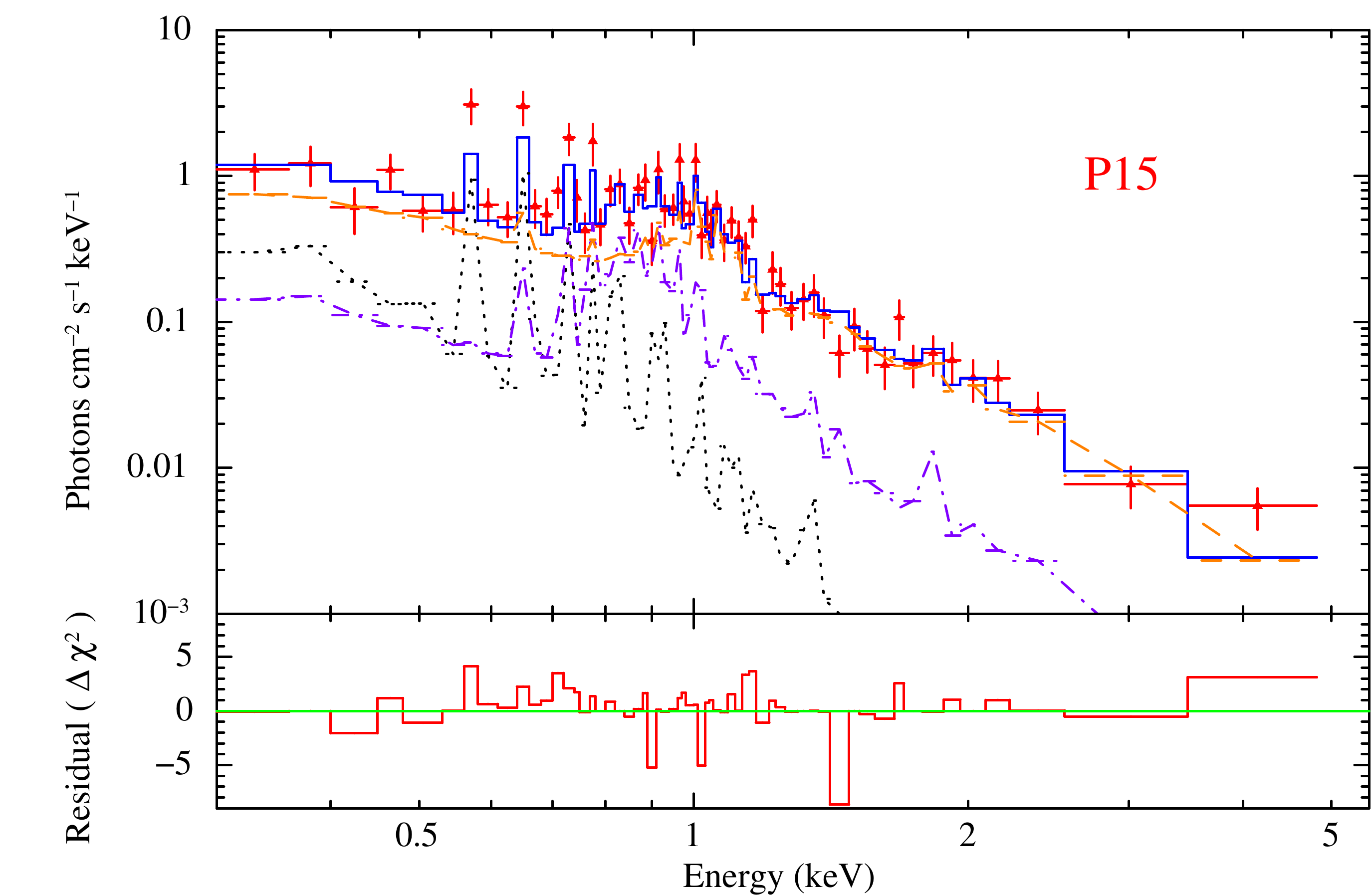}}

\caption{Spectral fitting to the {\it AstroSat}/SXT data  have been shown for the time intervals as given in Figure~\ref{fig:lc} and Table~\ref{tab:fit_params}. In each of the figures (`a' to `s'), the upper panel shows the data with red triangles. The best-fit 3-T \apec\ model is shown with a solid blue line. The black dotted line, purple dot-dashed line, and orange dashed lines shows the unfolded 1-T, 2-T, and 3-T temperature component to the model, respectively. The bottom panel shows the residual in the unit of $\Delta$\chisq. The top-right corner of the upper panel in each plot shows the segment number (i.e., P01 to P19; as mentioned in Table~\ref{tab:fit_params}). Here flare F1, F2, and F3 corresponds to the segments P03--P07, P08--P12, P13--P18, respectively. Whereas the segments P01--P02 and P19 indicate the `Pre-Flare' and the `Post-Flare' Spectra, respectively.}
\label{fig:spectra}
\end{figure*}

\begin{figure*}
\setcounter{figure}{1}
\addtocounter{subfigure}{15}
\centering
\subfloat[]{\includegraphics[height=6.0cm, angle=-90]{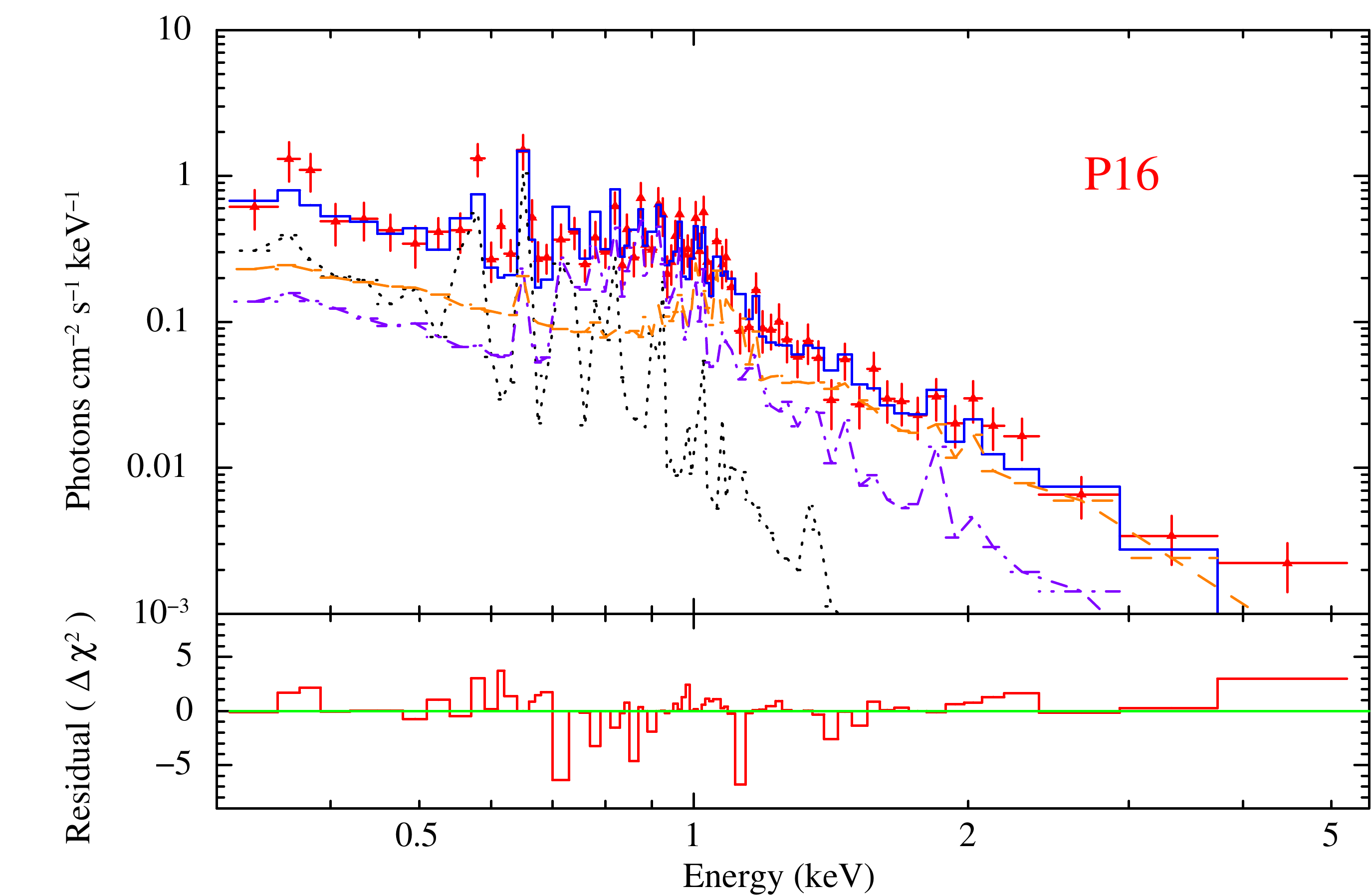}}
\subfloat[]{\includegraphics[height=6.0cm, angle=-90]{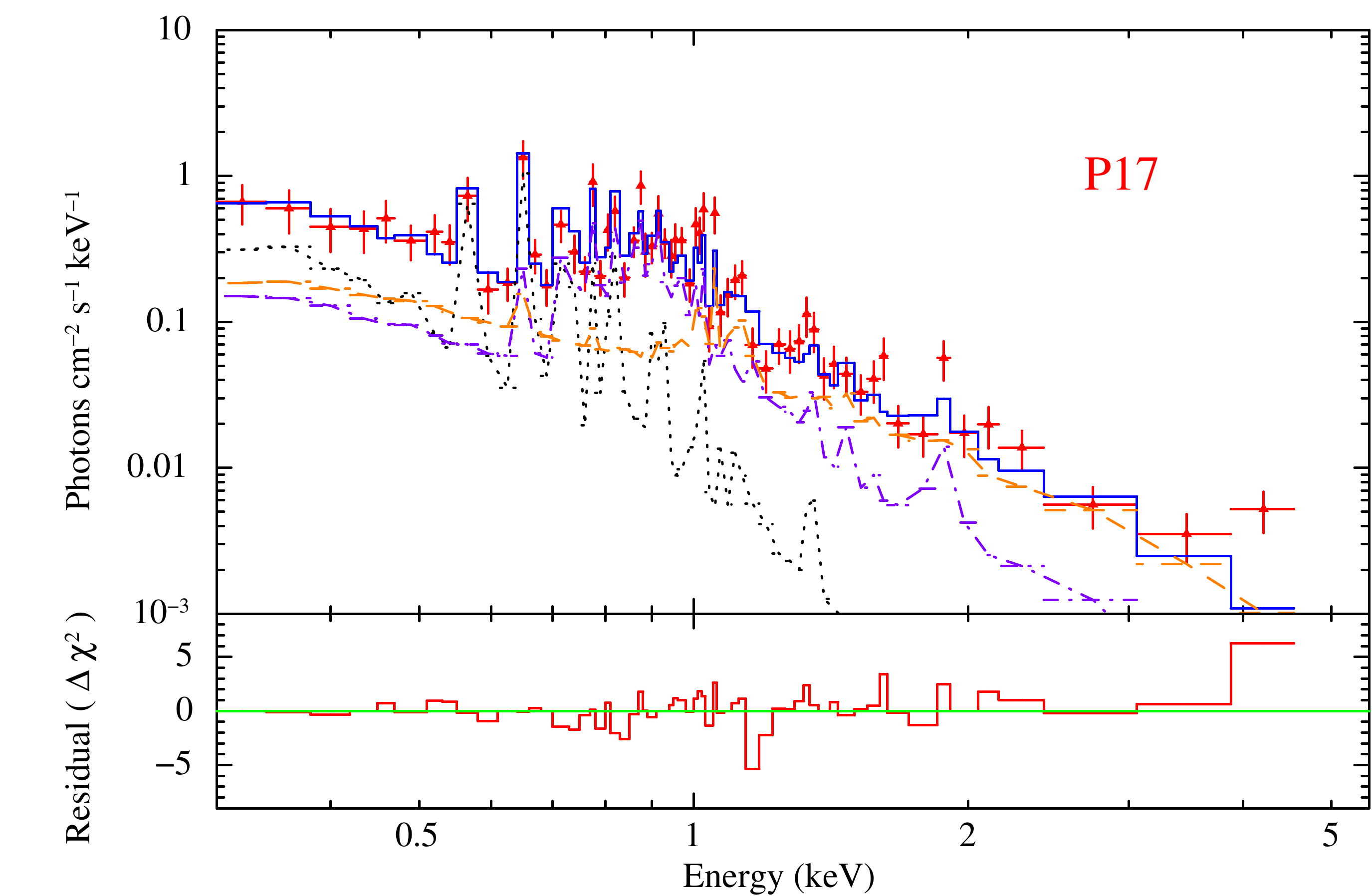}}
\\[-4.2mm]
\subfloat[]{\includegraphics[height=6.0cm, angle=-90]{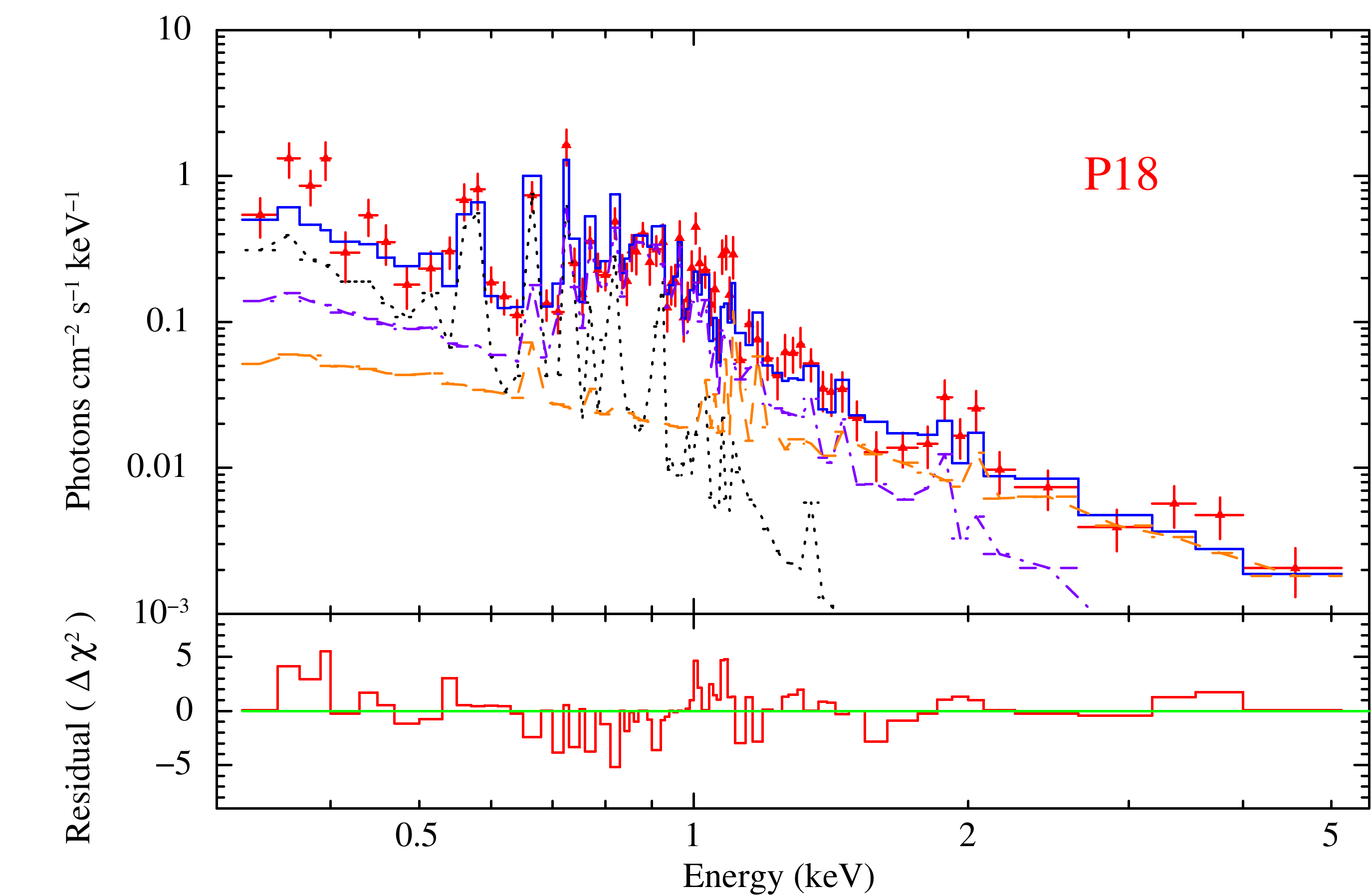}}
\subfloat[]{\includegraphics[height=6.0cm, angle=-90]{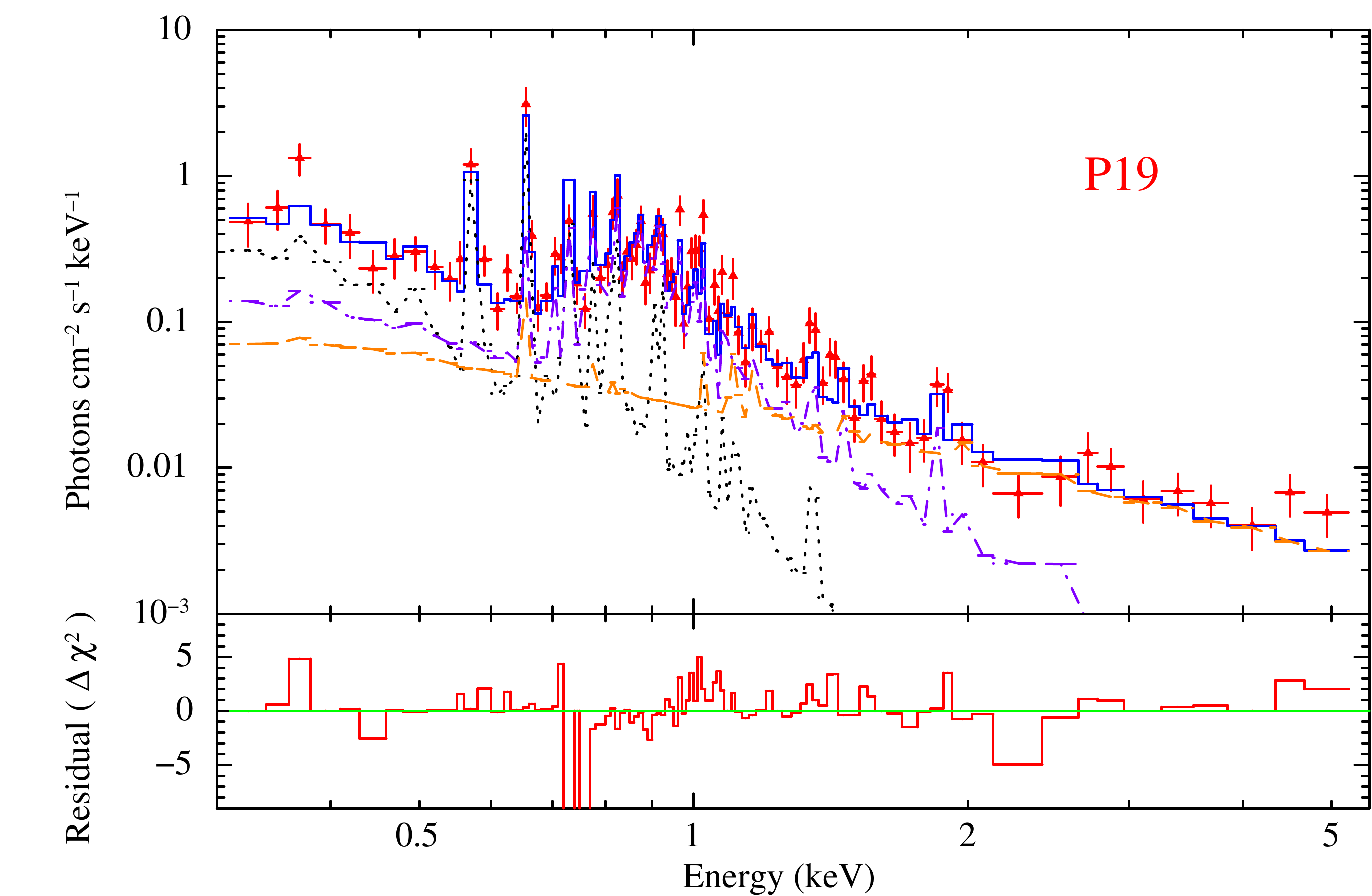}}

\caption{Continued.}
\end{figure*}

\subsection{Spectral Analysis}\label{sec:spec}

In this section, we provide a detailed description of the X-ray spectral analysis of \eqp\ using {\it AstroSat}/SXT data. Since the spectral parameters evolve with time, to trace the changes in the source during the observation, we performed time-resolved spectroscopy. Considering the data-gap in the SXT light curve,  in order to have approximately equal total counts in all the segments, we have divided the SXT light curves into nineteen segments (P01 to P19), which includes five (P03--P07), five (P08--P12), and six (P13--P18) segments for the flares F1, F2, and F3, respectively, whereas two segments (P01 and P02) and one (P19) segment are for `Pre-Flare' and `Post-Flare'. All these segments are shown with grey shaded regions, separated by vertical dotted lines in Figure~\ref{fig:lc}. Spectrum for each segment was extracted for further analysis. 

\subsubsection{Post-Flare Spectra}\label{sec:spec_pf}

During the $\sim$140 ks of {\it AstroSat} observation of \eqp, no distinct quiescent state was observed. For preliminary analysis, therefore, as a starting point, we chose a segment with the lowest mean count rate as the `proxy' of the quiescent phase. Although the segments P13 and P14 have similar lowest-mean count rates, we chose `P14' as the `Post-Flare' region as the data in this segment might not have been contaminated due to the flare F3. The coronal parameters of the `Post-Flare' phase were derived by fitting the spectrum with single (1-T), two (2-T), and three (3-T) temperatures Astrophysical Plasma Emission Code   \citep[\apec; see][]{Smith-01-ApJ-96} as implemented for collisionally ionized plasma. The global abundances (Z) and interstellar H{\sc i} column density (\nh) were left as free parameters. None of the plasma models (1-T, 2-T, or 3-T) with solar abundances (\zsun) was formally acceptable due to the large values of \chisq. Although the 2-T model with sub-solar abundances was found to have a significantly better fit than that of the 1-T plasma model, the fitting is still unacceptable for the `Post-Flare' spectrum as the value of reduced \chisq\ was 2.82 for 45 DOF. A 3-T plasma model improved the fitting significantly for the Post-Flare segment. The best-fit temperatures are derived to be 0.25$\pm$0.02, 0.79$\pm$0.04, and $>$1.09 keV for the Post-Flare segment. The corresponding global abundances were derived to be 0.37$\pm$0.03, 0.41$\pm$0.04, and $>$0.03 \zsun. The unabsorbed X-ray Luminosity (L$_{X}$) were calculated in 0.3--7 keV range using the {\tt CFLUX} model for a distance of 6.260$\pm$0.003 pc \citep{Bailer-Jones-18-AJ-6}. The L$_{X}$ is derived to be 2.7$\times$10$^{30}$ erg s$^{-1}$, which is $\sim$17 times more luminous than the equivalent quiescent X-ray luminosity of \eqp, estimated from the \chandra/HETG observations \citep{Liefke-08-A+A-2}.
This result indicates that the `Post-Flare' is not the actual quiescent state of \eqp. In order to compute the equivalent quiescent luminosity, we have converted the flux from 2--25 \AA\  \chandra/HETG band to 0.3--7 keV energy band using {\sc webpimms}\footnote{\href{https://heasarc.gsfc.nasa.gov/cgi-bin/Tools/w3pimms/w3pimms.pl}{https://heasarc.gsfc.nasa.gov/cgi-bin/Tools/w3pimms/w3pimms.pl}}. For this conversion, we have considered a 3-T \apec\ model of \eqp\ with the same parameters as derived in this section.  

\subsubsection{Flare Spectra}\label{sec:spec_flare}

\begin{figure*}
\centering
\includegraphics[width=16.5cm, angle=0]{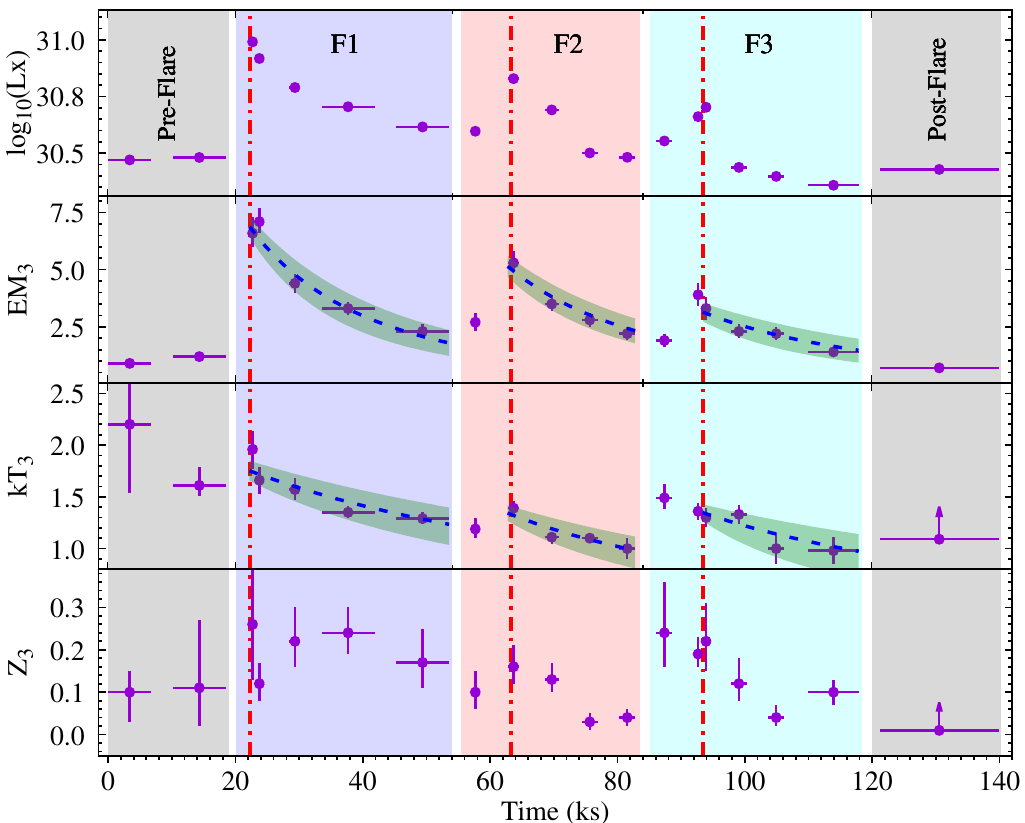}
\caption{Evolution of spectral parameters of \eqp. From top to bottom panels, the logarithmic values of X-ray luminosities as derived in 0.3--7 keV band, the emission measure ($EM_3$) associated to flaring plasma,  corresponding plasma temperature ($kT_3$), and corresponding global abundance ($Z_3$) have been shown. 
  The blue, orange, and cyan shaded regions indicate the duration of flares F1, F2, and F3. The `Pre-Flare' and `Post-Flare' segments have been shown with the grey shaded region. The Horizontal bars give the time range over which spectra were extracted, and the vertical bars show a 68\% confidence interval of the parameters. 
  The blue dashed line shows the best-fit of Equation~\ref{eq:qs_em} and  \ref{eq:qs_kT} to the $EM_3$ and $kT_3$, respectively during the decay phase of each flare. The green shaded region shows the 1$\sigma$ uncertainty in the fitting. The red dot-dashed vertical lines show the flare peak time t$_{0}$ for each flare as mentioned in Equations~\ref{eq:qs_em} and \ref{eq:qs_kT}, and described in Section~\ref{sec:loop}. The best-fit quasi-static decay times estimated from each flare have been given in Table~\ref{tab:loop_params}.
}
\label{fig:spec_params}
\end{figure*}

As mentioned in Section~\ref{sec:lc}, there are three flaring events detected during the SXT observation of \eqp. Spectra corresponding to all the segments during the flaring events were extracted and fitted with 1-T, 2-T, and 3-T plasma models. Among these three models, the 3-T plasma model was found to be acceptable for fitting the spectra of all the segments of flares F1, F2, and F3. In the beginning, the fitting was carried out by keeping all parameters such as three temperatures, normalizations, \nh, and abundances free. While fitting, though \nh\ was a free parameter, its value was found to be constant (within 1$\sigma$ uncertainty level). Considering this, we fixed \nh\ to its average value of 2.32$\times$10$^{20}$ atoms cm$^{-2}$. This average derived value of \nh\ is marginally less than that estimated in a cone of 0.1 degrees in the direction of \eqp\ using the latest H{\sc i} Pi survey \citep[3.95$\times$10$^{20}$ atoms cm$^{-1}$;][]{HI4PI-Collaboration.-16-A+A-1}.
  Since the distance of \eqp\ is minimal compared to the thin disk scale height \citep[$\sim$300~pc;][]{GilmoreG-83-MNRAS-3}, it is unlikely that the large estimated value of the \nh\ to be contributed from the interstellar medium between the Sun and \eqp. Instead, it is more likely that the estimated column density is due to the locally originated gas from \eqp.
The first two temperatures, corresponding EMs, and corresponding abundances were found to be constant within the 1$\sigma$ uncertainty level. The average values of the first two temperatures over all the segments are 0.24$\pm$0.04 and 0.81$\pm$0.06 keV, whereas the corresponding EMs are derived to be 6.98$\pm$0.07 $\times$10$^{52}$ and 6.04$\pm$0.08 $\times$10$^{52}$ cm$^{-3}$, respectively. The abundances corresponding to the first two temperatures are also found to be equal $\sim$0.4 within 1$\sigma$ uncertainty. These two temperatures, EMs, and abundances are found to be similar to that of the first two temperatures of the Post-Flare phase. Therefore, for the further spectral fitting of the flare-segments, we fixed the first two temperatures, corresponding EMs, and abundances at the average values. Only the third temperature (kT$_{3}$), corresponding EM (EM$_{3}$), and abundance (Z$_{3}$) are allowed to vary while fitting the flaring segments and identified as flare components. The time evolution of derived spectral parameters during the flare F1, F2, and F3 are shown in Figure~\ref{fig:spec_params} using blue, orange, and cyan shaded regions, respectively. The other segments (including the Post-Flare segment) are shown with the grey shaded region. The derived spectral parameters from fitting the spectra from all the segments with a 3-T plasma model are given in Table~\ref{tab:fit_params}.

The kT$_{3}$, EM$_{3}$, Z$_{3}$, and L$_{X}$ were found to vary during the flares. The peak values of abundances for flares F1, F2, and F3 were derived to be $\sim$0.26, $\sim$0.16, and $\sim$0.24 \zsun, which were 9, 5, and 8 times more than that of the post-flaring region. The derived peak flare temperatures for F1, F2, and F3 are 2, 1.4, and 1.5 times more than the lowest observed third thermal component. The EM$_{3}$ followed the flare light curve and peaked at a value of 7.1 \E{53}, 5.3 \E{53}, and 3.9 \E{53} cm$^{-3}$ for the flares F1, F2, and F3, which are approximately 10, 8, and 6 times than that of the Post-Flare region, respectively. The peak values of log$_{10}$(\lx) in 0.3--7 keV energy band during flares F1, F2, and F3 are derived to be 30.99$\pm$0.01, 30.83$\pm$0.01, and 30.70$\pm$0.02 (\lx\ is in the unit of erg s$^{-1}$). This shows that the flares F1, F2, and F3 are approximately 4, 3, and 2 times more luminous than that of the Post-Flare regions.

\begin{table*}
\centering
\begin{threeparttable}

\tabcolsep=0.43cm


\caption{Best-fit spectral parameters of \eqp\ obtained from the time-resolved spectroscopy of {\it AstroSat}/SXT observation.}\label{tab:fit_params}
\begin{tabular}{ccccccccccccccccc}
\toprule
$\mathbf{^{}Flares}$&
$\mathbf{^{}Segments}$&
$\mathbf{^{}Time range}$&
$\mathbf{^{}kT_3}$&
$\mathbf{^{}EM_3}$&
$\mathbf{^{}Z_3}$&
$\mathbf{^{}log_{10}(Lx)}$&
$\mathbf{^{}}$\textbf{\chisq (DOF)}\\

&
&
(ks)&
(keV)&
($10^{53}$ cm$^{-3}$)&
(\zsun)&
(cgs unit)&
\\

 \midrule
  
Pre-Flare
 	&P01	&   T0+0   	    --	T0+6.79	    &		$2.20_{-0.66}^{+0.87}$	&		$0.9_{-0.1}^{+0.1}$	&		$0.10_{-0.07}^{+0.05}$	&		$30.47_{-0.01}^{+0.01}$	&		1.046~(87)		\\
    &P02	&   T0+10.14   	--	T0+18.48	&		$1.61_{-0.10}^{+0.18}$	&		$1.2_{-0.2}^{+0.2}$	&		$0.11_{-0.09}^{+0.16}$	&		$30.48_{-0.01}^{+0.01}$	&		1.119~(119)		\\
\\
    F1	&P03	&   T0+22.21   	--	T0+23.16	&		$1.96_{-0.19}^{+0.18}$	&		$6.6_{-0.6}^{+0.7}$	&		$0.26_{-0.13}^{+0.14}$	&		$30.99_{-0.01}^{+0.01}$	&		1.048~(87)		\\
	&P04	&   T0+23.16   	--	T0+24.33	&		$1.66_{-0.13}^{+0.13}$	&		$7.1_{-0.5}^{+0.6}$	&		$0.12_{-0.04}^{+0.05}$	&		$30.92_{-0.01}^{+0.01}$	&		0.945~(90)		\\
	&P05	&   T0+28.48   	--	T0+30.18	&		$1.57_{-0.10}^{+0.11}$	&		$4.4_{-0.4}^{+0.4}$	&		$0.22_{-0.06}^{+0.08}$	&		$30.79_{-0.01}^{+0.01}$	&		1.028~(95)		\\
	&P06	&   T0+33.52   	--	T0+41.87	&		$1.35_{-0.05}^{+0.06}$	&		$3.3_{-0.3}^{+0.3}$	&		$0.24_{-0.05}^{+0.06}$	&		$30.70_{-0.01}^{+0.01}$	&		1.094~(115)		\\
	&P07	&   T0+45.21   	--	T0+53.56	&		$1.29_{-0.05}^{+0.06}$	&		$2.3_{-0.3}^{+0.3}$	&		$0.17_{-0.06}^{+0.08}$	&		$30.62_{-0.01}^{+0.01}$	&		1.054~(94)		\\
\\                                                                                                  
F2  &P08	&   T0+56.90   	--	T0+58.46	&		$1.19_{-0.09}^{+0.11}$	&		$2.7_{-0.4}^{+0.4}$	&		$0.10_{-0.04}^{+0.05}$	&		$30.60_{-0.02}^{+0.02}$	&		1.043~(59)		\\
	&P09	&   T0+62.75   	--	T0+64.61	&		$1.39_{-0.08}^{+0.07}$	&		$5.3_{-0.4}^{+0.5}$	&		$0.16_{-0.04}^{+0.05}$	&		$30.83_{-0.01}^{+0.01}$	&		1.174~(110)		\\
	&P10	&   T0+68.59   	--	T0+70.76	&		$1.11_{-0.07}^{+0.05}$	&		$3.5_{-0.3}^{+0.4}$	&		$0.13_{-0.03}^{+0.04}$	&		$30.69_{-0.01}^{+0.01}$	&		1.310~(92)		\\
    &P11	&   T0+74.44   	--	T0+76.88	&		$1.10_{-0.05}^{+0.05}$	&		$2.8_{-0.3}^{+0.3}$	&		$0.03_{-0.02}^{+0.02}$	&		$30.50_{-0.02}^{+0.01}$	&		1.151~(68)		\\
    &P12	&   T0+80.29   	--	T0+82.79	&		$1.00_{-0.10}^{+0.10}$	&		$2.2_{-0.3}^{+0.3}$	&		$0.04_{-0.02}^{+0.02}$	&		$30.48_{-0.02}^{+0.01}$	&		1.347~(69)		\\
\\                                                                                                  
F3  &P13	&   T0+86.13   	--	T0+88.64	&		$1.49_{-0.11}^{+0.13}$	&		$1.9_{-0.3}^{+0.3}$	&		$0.24_{-0.08}^{+0.12}$	&		$30.55_{-0.01}^{+0.01}$	&		1.056~(83)		\\
	&P14	&   T0+91.98   	--	T0+93.35	&		$1.36_{-0.08}^{+0.08}$	&		$3.9_{-0.5}^{+0.5}$	&		$0.19_{-0.03}^{+0.04}$	&		$30.66_{-0.02}^{+0.02}$	&		0.992~(60)		\\
	&P15	&   T0+93.35   	--	T0+94.48	&		$1.30_{-0.09}^{+0.09}$	&		$3.3_{-0.5}^{+0.5}$	&		$0.22_{-0.07}^{+0.09}$	&		$30.70_{-0.02}^{+0.02}$	&		1.248~(57)		\\
    &P16	&   T0+97.82   	--	T0+100.33	&		$1.33_{-0.09}^{+0.09}$	&		$2.3_{-0.3}^{+0.3}$	&		$0.12_{-0.04}^{+0.06}$	&		$30.44_{-0.02}^{+0.02}$	&		1.042~(68)		\\
    &P17	&   T0+103.72   --	T0+106.17	&		$1.00_{-0.15}^{+0.15}$	&		$2.2_{-0.3}^{+0.3}$	&		$0.04_{-0.02}^{+0.03}$	&		$30.40_{-0.02}^{+0.02}$	&		1.072~(59)		\\
    &P18	&   T0+109.99   --	T0+117.86	&		$0.98_{-0.13}^{+0.13}$	&		$1.4_{-0.2}^{+0.2}$	&		$0.10_{-0.03}^{+0.03}$	&		$30.36_{-0.02}^{+0.02}$	&		1.412~(71)		\\
\\  
Post-Flare
    &P19	&   T0+121.20   --	T0+139.90	&		$> 1.09$	            &		$0.7_{-0.1}^{+0.1}$	&		$>$0.03 	  &		$30.43_{-0.01}^{+0.01}$	&		1.208~(92)		\\

 \bottomrule
\end{tabular}

\normalsize

\end{threeparttable}
\end{table*}

\section{Discussion and Conclusions}
\label{sec:discussion}
In this section, we have presented the loop modelling, energetics, coronal properties, and mass loss due to CMEs associated with the X-ray flares as observed on \eqp. We have also discussed our results in the light of present understanding.

\subsection{Loop~Modeling}\label{sec:loop}

Stellar flares cannot be spatially resolved, although it is possible to infer the physical size and structure of the flares using the flare loop models. Based on quasi-static radiative and conductive cooling during the decay, \cite{van-den-OordG-89-A+A-1} suggested an approach to model a loop.  According to this model, the ratio of the radiative cooling time ($\tau_{rad}$) and the conductive cooling time ($\tau_{cond}$) remains constant during the decay phase of the flare. The two timescales are expressed by the following formula.
\begin{equation} 
  \centering
\tau_{rad} = \frac{3 k T}{n_e P(T)} ~~~~{\rm and} ~~~\tau_{cond} = \frac{3 n_e k L^2}{\kappa T^{5/2}}
\label{eq:trad}
\end{equation}
\noindent
where $T$ is the loop temperature,  $n_e$ is electron density, $k$ is Boltzmann constant, $\kappa$ is the thermal conductivity, and $P(T)$ is the combined plasma emissivity per unit emission measure of X-ray emission lines and bremsstrahlung continuum spectrum, and is given by $P(T) = 10^{-24.73} {T}^{1/4}$ erg cm$^{3}$ s$^{-1}$. Both decay times ($\tau_{rad}$ and $\tau_{cond}$) depend on the loop-length $L$, while $\tau_{cond}$ explicitly through the $L^2$ term, $\tau_{rad}$ implicitly through the density dependence.

Assuming the shape of the flaring loops to be semicircular with a constant cross-section throughout the flare, the quasi-static decay time scale can be determined using the Equation~26 of \citep[][]{van-den-OordG-89-A+A-1} as
\begin{equation}
  \centering
  EM(t) = EM(t_0) \left(1+ \frac{t - t_0}{3\tau_{qs,EM}}\right)^{-26/7}
  \label{eq:qs_em}
\end{equation}
\begin{equation}
  \centering
  kT(t) = kT(t_0) \left(1+ \frac{t - t_0}{3\tau_{qs,kT}}\right)^{-8/7}
  \label{eq:qs_kT}
\end{equation}
Where $t_0$ is an arbitrary initial epoch. In the present case, $t_0$ is considered as the epoch at which X-ray luminosity Lx peaked for each flare.  Whereas $\tau_{qs,EM}$ and  $\tau_{qs,kT}$ are the quasi-static cooling timescales. The emission measure $EM(t)$ and  temperature $kT(t)$ are the time-dependent parameters.

In order to estimated the quasi-static cooling timescales, we have fitted Equation~\ref{eq:qs_em} to the EM$_{3}$ data and Equation~\ref{eq:qs_kT} to the kT$_{3}$ data that were  derived from the spectral fitting to the {\it AstroSat}/SXT data. The decay phase of the flares F1, F2, and F3 contain five (P03 - P07), four (P09 - P12), and four (P15 - P18) data points, respectively. We have used the orthogonal distance regression technique to get the best fit considering uncertainties in both x-y directions.
 As a result, we estimate $\tau_{qs,EM}$ values for flares F1, F2, and F3 to be 22$\pm$4, 22$\pm$5, and 28$\pm$10 ks with the best-fit reduced \chisq\ (DOF) values of 0.7 (4), 0.8 (3), and 0.7 (3), respectively. Whereas the $\tau_{qs,kT}$ is estimated to be 25$\pm$7, 18$\pm$6, and 21$\pm$11 ks with the best-fit reduced \chisq\ (DOF) values of 0.6 (4), 0.7 (3), and 0.6 (3).
In Equations~\ref{eq:qs_em} and \ref{eq:qs_kT}, two timescales ($\tau_{qs,EM}$ and  $\tau_{qs,kT}$) are expected to be same when the flare cools quasi-statically. For all three flares F1, F2, and F3, we confirm that both the time scales are indeed same (within the errors) i.e. $\tau_{qs,EM}$ $\approx$  $\tau_{qs,kT}$. 
Therefore, in the following analysis, we use $\tau_{qs,EM}$ as the cooling timescale (= $\tau_{qs}$) for all three flares, since it is better constrained than $\tau_{qs,kT}$.
The evaluated values of $\tau_{qs}$ for the flares F1, F2, and F3 are 22$\pm$4, 22$\pm$5, 28$\pm$10 ks, respectively. 

Using the fitted values of $\tau_{qs}$, we derived three geometric parameters i.e flare loop length (L),  loop aspect ratio (a), and the loop-density (n$_e$) using the following three equations of the quasi-static cooling model \citep[][]{van-den-OordG-89-A+A-1, TsuboiY-00-ApJ-4},
\begin{equation}
  L ~(cm) = \rm{R}_{\odot}~{ \left(\frac{\tau_{qs}}{10~{\rm ks}}\right)} ~ \left(\frac{kT(t_0)}{{\rm keV}}\right)^{7/8} 
  \label{eq:L}
\end{equation}

\begin{equation}
  n_e ~(cm^{-3}) = 4.4\times10^{10} ~   { \left(\frac{\tau_{qs}}{10~{\rm ks}}\right)}^{-1} ~  \left(\frac{kT(t_0)}{{\rm keV}}\right)^{4/3}
  \label{eq:ne}
\end{equation}

\begin{equation}
  a = 1.38 ~ { \left(\frac{\tau_{qs}}{10~{\rm ks}}\right)}^{-1/2} ~  \left(\frac{kT(t_0)}{{\rm keV}}\right)^{-33/16} ~  \left(\frac{EM(t_0)}{{10^{54} \rm cm^{-2}}}\right)^{1/2} 
  \label{eq:a}
\end{equation}
Using the quasi-static cooling model, the flaring loop lengths for the flares F1, F2, and F3 are derived to be 2.5$\pm$0.5~\E{11}, 2.0$\pm$0.5~\E{11}, 2.5$\pm$0.9~\E{11} cm, respectively. Assuming the semi-circular geometry, the flaring loop heights (L/$\pi$) of flares F1, F2, and F3 are estimated as 2.9, 2.3, and 2.8 times than that of the stellar radius of the primary component of \eqp. 

From Equation~\ref{eq:ne}, we derived the density of the flaring loops of F1, F2, and F3 to be 4.2$\pm$0.8\E{10}, 3.0$\pm$0.7\E{10}, and 2.2$\pm$0.8\E{10} cm$^{-3}$, respectively. Using the values of $n_e$, we derived the maximum pressure (p) in the loop at the flare peak using the relation $p = 2 n_e~k ~T_\mathrm{max}$. The estimated values of $p$ for flares F1, F2, and F3 are 2.4$\pm$0.5~\E{2}, 1.3$\pm$0.3~\E{2}, and 0.9$\pm$0.4~\E{2} dyne cm$^{-2}$, respectively. The flaring loop volume ($V$) and the minimum magnetic field ($B$) to confine the flaring plasma can be derived by using the following relation :
\begin{equation}
  V~~({\rm cm^3}) = \frac{EM_{3,max}}{n_e^{2}}  ;~~~~~B~~({\rm G}) = \sqrt{8\pi ~p}
 \label{eq:V-B}
\end{equation}
The estimated values of  $V$ during the flares F1, F2, and F3 are 4$\pm$2~\E{32}, 6$\pm$3~\E{32}, and 7$\pm$5~\E{32} cm$^3$, whereas the values of the minimum magnetic field $B$ are derived to be 69--85, 48--64, 39--57 G, respectively. The loop parameters for all three flares are given in Table~\ref{tab:loop_params}. 
Using Equation~\ref{eq:a}, we have derived the diameter-to-length (aspect) ratio (=a) for flares F1, F2, and F3 of \eqp\ to be 0.24$\pm$0.04, 0.37$\pm$0.06, 0.28$\pm$0.07. 
These values of aspect ratios are within the range of 0.1 -- 0.3 for typical solar and stellar coronal loops \citep[e.g.][]{Shimizu-95-PASJ-3, SciortinoS-99-A+A}.

\begin{table*}
\centering
\begin{threeparttable}

\tabcolsep=0.95cm
\caption{Loop Parameters derived for flares F1, F2, and F3 of \eqp}\label{tab:loop_params}
\begin{tabular}{lcccccccccccccccc}
\toprule
\textbf{Sl.} &\textbf{Parameters}                            &   \textbf{Flare F1}          &   \textbf{Flare F2}         &   \textbf{Flare F3}       \\
\hline
1  &   $\tau_\mathrm{r}$  ($\mathrm{ks}$)                    &   --                         &   3.4 $\pm$  0.8            & 11  $\pm$    2    \\
2  &   $\tau_\mathrm{d}$  ($\mathrm{ks}$)      &   1.6 $\pm$ 0.4 , 24 $\pm$  5              &   3.0 $\pm$  0.6            & 3.1 $\pm$    0.4           \\
3  &   $\tau_\mathrm{qs,EM}$ ($\mathrm{ks}$)                 &   22 $\pm$ 4                 &   22 $\pm$5                 & 28   $\pm$ 10            \\
4  &   $\tau_\mathrm{qs,kT}$ ($\mathrm{ks}$)                 &   25 $\pm$ 7                 &   18 $\pm$6                 & 21   $\pm$ 11            \\
5  &   $EM(t_\mathrm{0}$) ($10^{53}~\mathrm{cm^{-3}}$)       &   6.9 $\pm$ 0.4              &   5.2 $\pm$0.5              & 3.3   $\pm$ 0.5            \\
6  &   $kT(t_\mathrm{0}$) ($\mathrm{keV}$)                   &   1.75 $\pm$ 0.09            &   1.34 $\pm$0.08            & 1.30   $\pm$ 0.09            \\
7  &   ${\rm log}_{10}(L_\mathrm{X, max}$) (in~cgs)          &   30.99 $\pm$ 0.01           &   30.83 $\pm$ 0.01          & 30.70 $\pm$ 0.02   \\
8  &   $L$ ($10^{11}~\mathrm{cm})$                           &   2.5 $\pm$ 0.5              &   2.0 $\pm$ 0.5             & 2.5 $\pm$ 0.9        \\
9  &   a                                                     &   0.24$\pm$0.04              &   0.37$\pm$0.06             & 0.28$\pm$0.07         \\
10  &   $n_\mathrm{e}$ ($10^{10}~\mathrm{cm^{-3}})$          &   4.2 $\pm$ 0.8              &   3.0 $\pm$ 0.7             & 2.2 $\pm$ 0.8          \\
11  &   $p$ ($10^{2}~\mathrm{dyn~cm^{-2}})$                  &   2.4$\pm$0.5                &   1.3$\pm$0.3               & 0.9$\pm$0.4           \\
12  &   $V$ ($10^{32}~\mathrm{cm^{3}})$                      &   4$\pm$2                    &   6$\pm$3                   & 7$\pm$5  \\
13  &   $B$ ($\mathrm{G})$                                   &    69 -- 85                  &   48 -- 64                  &  39 -- 57    \\
14  &   HR$_\mathrm{V}$ ($\mathrm{10^{-2}~erg~s^{-1}~cm^{-3}})$&  1.9$\pm$0.7               &   1.1$\pm$0.5               & 0.6$\pm$0.5          \\
15  &   HR ($10^{30}~\mathrm{erg~s^{-1}})$                   &   7.2$\pm$0.5                &   6.5$\pm$0.7               & 4.2$\pm$0.7              \\
16  &   $E_\mathrm{X}$ ($10^{34}~\mathrm{erg})$              &   $\sim$12.7                 &   $\sim$5.7                 & $\sim$2.4          \\
17  &   $E_\mathrm{H}$ ($10^{34}~\mathrm{erg})$              &   $\gsimeq$15.2              &   $\sim$4.2                 & $\gsimeq$1.3         \\
18  &   M$_{\rm CME}$ ($10^{18}~g)$                          &  $\sim$9.3                   &  $\sim$5.8                  &  $\sim$3.5  \\[0.5mm]
19  &   {v}$_{\rm esc}$ (km~s$^{-1}$)                        &  $\sim$5500                  &  $\sim$4600                 &  $\sim$3900 \\[0.5mm]
20  &   $E_{\rm KE,CME}$ ($10^{35}~\mathrm{erg})$            &  $\sim$14                    &  $\sim$2                    &  $\sim$3 \\
21  &   $B_\mathrm{tot}$ ($\mathrm{G})$                      &  106 -- 146                  &  58 -- 82                   &  59 -- 77   \\
\hline
\end{tabular} 
\begin{tablenotes}
\item  \textbf{Note.} 
\textbf{1, 2} -- e-folding rise- and decay-time as derived from the light curve; 
\textbf{3, 4} -- quasi-static decay times as derived by fitting Equations~\ref{eq:qs_em} and \ref{eq:qs_kT} (see Section~\ref{sec:loop}); 
\textbf{5} -- Emission Measure related to flaring plasma at the flare peak; 
\textbf{6} -- Temperature related to flaring plasma at the flare peak; 
\textbf{7} -- X-ray Luminosity at the flare peak, estimated in \mbox{0.3 -- 7 keV} energy band; 
\textbf{8} -- length of the flaring loops; 
\textbf{9} -- loop aspect ratio i.e. diameter to  length ratio of the loop; 
\textbf{10, 11} -- Estimated maximum electron density and  loop pressure at flare peak;
\textbf{12} -- loop volume of the flaring region.
\textbf{13}-- Minimum magnetic field. 
\textbf{14} -- Heating rate per unit volume; 
\textbf{15} -- Total Heating rate;
\textbf{16} -- X-ray energy estimated using trapezoidal  integration  of  the derived X-ray luminosity.
\textbf{17} -- Estimated energy during flare due to heating of the stellar corona.
\textbf{18} -- Ejected coronal mass during the flaring events. 
\textbf{19} -- Outward escape velocity of ejected coronal mass. 
\textbf{20} -- Kinetic energy of the ejected coronal mass. 
\textbf{21} -- Total magnetic field required to produce the flare. 
(See the text for a detailed description).
\end{tablenotes}
\end{threeparttable}
\end{table*}

\subsection{Energetics} \label{sec:energetics}

A detailed assessment of the energy balance of the flares in the present case is not possible due to the lack of multi-wavelength coverage and velocity information, which could help to assess the plasma kinetic energy. Using simple trapezoidal integration of the instantaneous X-ray luminosity, we estimated the total energy radiated in X-rays (E$_{X}$) during flares F1, F2, and F3 are $\sim$12.7~\E{34}, $\sim$5.7~\E{34}, and $\sim$2.4~\E{34} erg. Over 38, 26, and 36 ks duration, the energy released in X-ray are equivalent to $\sim$13, $\sim$6, and $\sim$3 seconds of the bolometric energy output of the star (considering total bolometric energy of the primary and secondary components). With the assumption that the flaring loop is in the steady-state, using the scaling law of \cite{Rosner-78-ApJ-1}, we can estimate the heating rate per unit volume ($HR_{V} = \frac{d H}{d V d t} \simeq 10^5 ~ p^{7/6} ~ L^{-5/6}$) at the peak of the flare as 1.9$\pm$0.7, 1.1$\pm$0.5, 0.6$\pm$0.5 $\rm erg ~ cm^{-3} ~ s^{-1}$ for flares F1, F2, and F3, respectively. The total heating rate ($HR = \frac{d H}{dt} \simeq \frac{d H}{d V d t} \times V$) at the peak of the flare, was derived to be 7.2$\pm$0.5~\E{30} for F1, 6.5$\pm$0.7~\E{30} for F2, and  4.2$\pm$0.7~\E{30} \ergs\ for F3, which are 74\%, 96\%, and 84\% of the flare maximum luminosity, respectively. This result suggests that X-ray radiation is one of the major energy loss terms during the flare. If we assume that the heating rate is constant for the initial rise phase and then decays exponentially, with an e-folding decay time similar to that of the light-curve, then the total energy radiated during the flare is given by  $E_{\rm H} \sim \frac{d H}{dt} \times (\tau_r + \tau_d )$. The estimated values of total energy radiated during the flare are found to be $\gsimeq$~15.2~\E{34}, $\sim$4.2~\E{34}, $\gsimeq$1.3~\E{34}~erg. These values are of the same order as the estimated total energy radiated in the X-ray. All three flares are found to emit a large amount of flare energies ($>$\Pten{33} erg), and hence can be classified as `superflares'.

\subsection{Coronal Properties} \label{sec:corona_prop}

The X-ray luminosity during flares F1, F2, and F3 reaches up to $\sim$9.8~\E{30}, $\sim$6.8~\E{30}, and  $\sim$5.0~\E{30} \ergs, respectively. Among the previously reported flares on \eqp, the largest flare was observed with the \rosat\ HRI instrument \citep{KatsovaM-02-ASPC-1}. The \rosat\ flare had a peak \lx\ of 1.36~\E{30} \ergs\ in 0.1--2.4 keV energy range. To compute an equivalent luminosity in the 0.3--7 keV range, we considered 3-T \apec\ model as derived in this paper. Using {\sc webpimms}\footnote{\href{https://heasarc.gsfc.nasa.gov/cgi-bin/Tools/w3pimms/w3pimms.pl}{https://heasarc.gsfc.nasa.gov/cgi-bin/Tools/w3pimms/w3pimms.pl}}, the estimated values of the peak \lx\ of the largest flare with \rosat\ was 1.56~\E{30} \ergs. Therefore, all three flares observed during the {\it AstroSat} observation are among the most luminous X-ray flares ever observed on \eqp. Using \chandra\ observations, \cite{Liefke-08-A+A-2} have estimated the log$_{10}$(\lx/\lbol) of \eqp\ to be --3.23. During the flares presented here, the values of log$_{10}$(\lx/\lbol) were estimated to be  --3.00$\pm$0.01, --3.16$\pm$0.01, and --3.29$\pm$0.02 for flares F1, F2, and F3, respectively. This shows that the X-ray emission during the flaring events is not far from the `saturation' level \citep[][]{PizzolatoN-03-A+A-3}. The duration of the flares F1, F2, and F3 are found to be around 11, 7, and 10 hr.
We recognize the fact that due to the periodic data-gap and the typical uncertainty in SXT count rate of 0.05 \cts, there are possibility of several microflares during the \asts\ observation, however, the detected flare-durations are among the largest flare durations compared to the previous observations of \eqp, e.g. flare duration of 0.36 hr during 1994 \citep[from \rosat\ data;][]{KatsovaM-02-ASPC-1}, $\sim$3 hr during 2000 \citep[from \xmm\ data; ][]{RobradeJ-04-A+A},  $>$ 2 hr during 2006 \citep[from \chandra\ data;][]{Liefke-08-A+A-2}.

For the first time, we have carried out a detailed time-resolved spectral analysis of the flaring events on \eqp. Although earlier studies using \xmm\ and \chandra\ observations clearly resolved two stars of the binary system 5.$\!$\arcs8 apart, due to the large point spread function of \asts/SXT instrument, we could not identify the individual components. We found that the corona of the combined system of \eqp\ consists of three temperatures plasma. Two cooler temperatures are $\sim$2.9 and $\sim$9.2 MK and are not related to the flares. Other low-mass stars also show 2-T quiescent corona with similar temperatures, e.g., LO~Peg, V471~Tau, DG~CVn \citep[][]{Karmakar-16-MNRAS-8, Pandey-08-MNRAS-16, Osten-16-ApJ-3}. The third temperature component, which varies with the flares, is hotter than both the cooler temperature components and ranges from 12--26 MK. This value is of the similar order to those of the other superflares detected on AB~Dor \citep[][]{Maggio-00-A+A}, EV~Lac \citep[][]{Favata-00-A+A-3, Osten-10-ApJ-5},  II~Peg \citep[][]{Osten-07-ApJ-2},  and CC~Eri \citep[][]{Karmakar-17-ApJ-5}. 
Using the O{\sc vii} line in the \chandra/MEG observations of \eqp, \cite{Liefke-08-A+A-2} derived the corona temperatures to be $\approx$2 MK. This is comparable to the coolest temperature that we derived from our analysis. However, it is noteworthy that \cite{Liefke-08-A+A-2} also estimated the coronal temperature using other lines, i.e., Ne~{\sc ix}, Mg~{\sc xi}, and Si~{\sc xiii}, which is in the range of $\approx$2 -- 20 MK. Using the method of Differential Emission Measure, the authors also investigated the temperature of the individual stars \eqp~A and \eqp~B. Due to poor constraints, however, the authors could not derive the temperature of \eqp~A, though the temperature of \eqp~B was estimated to be 6.3--15.8 MK. These estimated values also contain the ranges of the derived temperature from our study.

 The global abundance related to flaring corona of \eqp\ shows variation up to $\sim$9 times than the minimum observed value. In the case of other superflares, the abundances were found to increase between 2--3 times than that of the quiescent level \citep{Favata-00-A+A-3, Favata-99-A+A-3, Maggio-00-A+A}. However, the variation of abundance ranges from none \citep[for II~Peg;][]{Osten-10-ApJ-5} to around 11 times \citep[for CC~Eri;][]{Karmakar-17-ApJ-5}.

Using quasi-static modelling, we have estimated the density of the flaring loop to be 4.2$\pm$0.8~\E{10}, 3.0$\pm$0.7~\E{10}, and 2.2$\pm$0.8~\E{10} cm$^{-3}$. \cite{Liefke-08-A+A-2} estimated the density of \eqp~A with the O{\sc vii} and Ne{\sc ix} lines to be 2--8~\E{10} and 15--38~\E{10} cm$^{-3}$, respectively. For \eqp~B, the estimated values were $<$9~\E{10} and $<$32~\E{10} cm$^{-3}$. Our analysis indicates that the estimated densities of the flaring coronal loops are of the same order.

\subsection{Coronal Mass Ejection} \label{sec:CME}
CMEs are the most energetic coronal phenomena that have been studied over few decades on active stars and the Sun. 
\cite{AarnioA-12-ApJ-1} and \cite{DrakeJ-13-ApJ-6} have estimated an empirical relationship between the solar flare X-ray energy and its associated CME mass as ${M}_{\rm CME}~~(g) = \mu ~ {E_{G}}^{ \gamma}$, 
where $\mu$ is a constant of proportionality, and $\gamma$ is the power law index, and $E_{G}$ is the X-ray energy in GOES 1--8 \AA\ band.  For magnetically active stars, \cite{DrakeJ-13-ApJ-6} estimated $\mu = 10^{-1.5 \mp 0.5} $ in cgs units and $\gamma = 0.59 \pm 0.02$. In order to estimate the CME mass, we have converted the 0.3--7 keV \asts\ flux to the 1--8 \AA\ GOES flux using {\sc webpimms}\footnote{\href{https://heasarc.gsfc.nasa.gov/cgi-bin/Tools/w3pimms/w3pimms.pl}{https://heasarc.gsfc.nasa.gov/cgi-bin/Tools/w3pimms/w3pimms.pl}} with a 3-T \apec\ model with the parameters of \eqp\ as derived in Section~\ref{sec:spec}. The estimated values of mean ejected mass $M_{\rm CME}$ for \eqp\ are found to be $\sim$9.3~\E{18},  $\sim$5.8~\E{18}, $\sim$3.5~\E{18}~g for flares F1, F2, and F3, respectively.
Recent studies have shown that the kinetic energy of the CME dominates the mechanical energy \citep[$E_{\rm CME}$ $\approx$  $E_{\rm KE,CME}$;][]{Emslie-12-ApJ-2}. 
For magnetically active stars, \cite{OstenR-15-ApJ-1} have shown evidence supporting an equipartition between the total radiated energy of the stellar flare and the mechanical energy of the associated CME, i.e. ${E_{\rm KE,CME}}~~({\rm erg}) = \frac{E_{X}}{\epsilon f_{X}}$. Here $f_{X}$  is the fraction of the bolometric radiated flare energy appropriate for the waveband in which the flare energy $E_{X}$ is being measured, and $\epsilon$ is a constant of proportionality which describes the relationship between bolometric radiated energy and CME kinetic energy \citep{Emslie-12-ApJ-2}. Adopting the values of $f_{X} \approx0.3$ for the soft X-ray band \citep{OstenR-15-ApJ-1}, and $\epsilon \approx0.3$ \citep{Emslie-12-ApJ-2}, the kinetic energies of the CME associated with the flares F1, F2, and F3 on \eqp\ are estimated to be $\sim$14~\E{35}, $\sim$2~\E{35}, and $\sim$3~\E{35} erg. The mean outward velocities of the associated CMEs on \eqp\ corresponding to the flares F1, F2, and F3 are estimated to be $\sim$5500, $\sim$4600, and $\sim$3900 km~s$^{-1}$, respectively. Previously on \eqp, using the radio and optical observations \cite{Crosley-18-ApJ-2} have estimated CME velocities associated with four flares ranging from 1315--1925 km~s$^{-1}$. High velocity plasma outflow were also detected on other M-dwarfs, e.g. AD~Leo \citep[1500--5800 km s$^{-1}$;][]{HoudebineE-90-A+A-1} and AU~Mic \citep[$\sim$1400 km s$^{-1}$;][]{KatsovaM-99-ApJ}.

\subsection{Magnetic Field Strength} \label{sec:magnetic}
In the present analysis, we have also made some relevant estimations of the magnetic field strength that would be required to accumulate the emitted energy and to keep the plasma confined in a stable magnetic loop configuration. If we consider that the energy released during the flare is indeed of magnetic origin, the total non-potential magnetic field $B_{\rm tot}$ involved in the flare can be estimated from the following relation:
\begin{center}
  \begin{equation}
    \label{eq:b}
    E_{\rm H} = {{(B_{\rm tot}^2-B^2)} \over 8 \pi}\times V
  \end{equation}
\end{center}
Assuming that the loop geometry does not change during the flare, $B_{\rm tot}$ is estimated to be 106--146, 58--82, 59--77 G for the flares F1, F2, and F3, respectively. Although our estimation of $B_{\rm tot}$ involves the loop volume $V$, this is not to indicate that the field fills up the whole volume. Rather it suggests that the flare energy is stored in the magnetic field configuration of an active region. 
Using  Zeeman Doppler Imaging map \cite{Morin-08-MNRAS-3} had estimated the photospheric magnetic field of 800 G for the primary and 1200 G for the secondary of \eqp. 

Although due to the large point spread function of \asts/SXT, the spatial information of the flares was not available, from the estimation of the magnetic field and loop length, we may provide a logical explanation about the origin of the flares. In the beginning, the possibility of the flares to have footpoints onto both of the stellar components, i.e., the primary and the secondary, can be rejected as the derived loop length is much less than the separation between the stellar components (5.43~\E{12} cm). Since \eqp\ is an M3.5+M4.5-dwarf binary, \eqp~A have a radiative interior with a thick convective envelope, whereas \eqp~B is a fully convective star. As a result,  the M3.5 star (\eqp~A) is supposed to be powered by solar-type `$\alpha-\Omega$' dynamo \citep[][]{RoaldC-97-MNRAS-2}, whereas the M4.5 star (\eqp~B) is supposed to be powered by a `turbulent' dynamo \citep[][]{Durney-93-SoPh-2}.  The `turbulent' dynamo is expected to produce small-scale magnetic fields. This is because there is no stable overshoot layer where the fields can be stored and amplified, and only small-scale magnetic regions should emerge uniformly to the surface. Since all three flares seem to be associated with the magnetic field of the order of 100 G, it is possible for either of the stars can provide such a magnetic field, and hence the flare may have occurred in any one of the components.

\section*{Acknowledgements}

This publication uses data from the \asts\ mission of ISRO, archived at the Indian Space Science Data Centre (ISSDC), for which the data was obtained from High Energy Astrophysics Science Archive Research Center (HEASARC), provided by NASA’s Goddard Space Flight Center. We thank the SXT Payload Operation Center at TIFR, Mumbai, for providing the necessary software tools.
 SK is grateful to Dr. Wm. Bruce Weaver and Dr. Craig Chester from Monterey Institute for Research in Astronomy (MIRA) for giving valuable inputs and discussion.
SK and JCP acknowledge the support of DST RFBR Indo-Russian Joint Research Grant reference INT/RUS/RFBR/P-167 and INT/RUS/RFBR/P-271. ISS acknowledges the support of the Ministry of Science and Higher Education of the Russian Federation under grant 075-15-2020-780 (N13.1902.21.0039).

\section*{Data availability}
The data underlying this article are available in the ISRO Science Data Archive (\href{https://astrobrowse.issdc.gov.in/astro_archive/archive/Home.jsp}{https://astrobrowse.issdc.gov.in/astro\_archive/archive/Home.jsp}).

\bibliographystyle{mnras}
\bibliography{SK_collections}
\end{document}